\documentclass[10pt,final,journal]{IEEEtran}
\usepackage[switch]{lineno} 
\usepackage{tabularx} 
\usepackage{booktabs}
\usepackage{multirow}
\usepackage{makecell}

\usepackage{stmaryrd}
\usepackage{graphicx}
\usepackage{amsthm} 
\usepackage{amsmath} 
\usepackage{bm}
\usepackage{graphicx}
\usepackage{epstopdf}
\usepackage{color}
\usepackage{float}
\usepackage[colorlinks=false,linkcolor=black,urlcolor=black,bookmarksopen=true, hidelinks]{hyperref}
\usepackage{bookmark}
\usepackage[flushleft]{threeparttable}
\usepackage{stackengine}
\usepackage{scalerel}
\usepackage{array}
\usepackage{stfloats}
\usepackage{lipsum}
\usepackage{mathtools}

\usepackage{xcolor,soul}

\usepackage{cite}

\ifCLASSINFOpdf
\else
\fi
\usepackage{amsfonts,amssymb}

\usepackage{amsmath}
\usepackage{algorithm}
\usepackage{algpseudocode}
\ifCLASSOPTIONcompsoc
 \usepackage[caption=false,font=normalsize,labelfont=sf,textfont=sf]{subfig}
\else
 \usepackage[caption=false,font=footnotesize]{subfig}
\fi
\begin{document}
%

\title{Efficient Rigid Body Localization based on Euclidean Distance Matrix Completion for AGV Positioning under Harsh Environment}

\author{Xinyuan~An,
        Xiaowei~Cui,
        Sihao~Zhao, \textit{Member, IEEE},
       Gang~Liu,
       Mingquan~Lu
 \thanks{This work was supported by the National Key R\&D Program of China under Grant No. 2021YFA0716603. \textit{(Corresponding author: Xiaowei Cui)}}
\thanks{X. An, X. Cui and G. Liu are with the Department of Electronic Engineering,
	Tsinghua University, Beijing 100084, China (e-mail: anxinyuan1983@163.com; cxw2005@tsinghua.edu.cn; liu\_gang@tsinghua.edu.cn).}
\thanks{S. Zhao is with NovAtel,  Autonomy \& Positioning division of
Hexagon, Calgary, T3K 2L5, Canada (e-mail: zsh01@tsinghua.org.cn).}
\thanks{M. Lu is with the Department of Electronic Engineering,
	Beijing National Research Center for Information Science and Technology, Tsinghua University, Beijing 100084, China. (e-mail: lumq@tsinghua.edu.cn).}

}

\markboth{}%
{Shell \MakeLowercase{\textit{et al.}}: Bare Demo of IEEEtran.cls for IEEE Journals}
%




\maketitle

\begin{abstract}
In real-world applications for automatic guided vehicle (AGV) navigation, the positioning system based on the time-of-flight (TOF) measurements between anchors and tags is confronted with the problem of insufficient measurements caused by blockages to radio signals or lasers, etc. Mounting multiple tags at different positions of the AGV to collect more TOFs is a feasible solution to tackle this difficulty. Vehicle localization by exploiting the measurements between multiple tags and anchors is a rigid body localization (RBL) problem, which estimates both the position and attitude of the vehicle. 
However, to the best of the authors' knowledge, the state-of-the-art solutions to the RBL problem do not deal with missing measurements, and thus will result in degraded localization availability and accuracy in harsh environments.
In this paper, different from these existing solutions for RBL, we model this problem as a sensor network localization problem with missing TOFs. 
To solve this problem, we propose a new  efficient RBL solution based on Euclidean distance matrix (EDM) completion, abbreviated as ERBL-EDMC. Firstly, we develop a method to determine the upper and lower bounds of the missing measurements to complete the EDM reliably, using the known relative positions between tags and the statistics of the TOF measurements. Then, based on the completed EDM, the global tag positions are obtained from a coarse estimation followed by a refinement step assisted with inter-tag distances. Finally, the optimal vehicle position and attitude are obtained iteratively based on the estimated tag positions from the previous step. Theoretical analysis and simulation results show that the proposed ERBL-EDMC method effectively solves the RBL problem with incomplete measurements. It obtains the optimal positioning results while maintaining low computational complexity compared with the existing RBL methods based on semi-definite relaxation (SDR). 

\end{abstract}


%
\IEEEpeerreviewmaketitle

\section{Introduction}\label{Introduction}
%
%
%
%
\IEEEPARstart {R}{E}AL-time, continuous and high-accuracy positioning is paramount for the reliable operation of automatic guided vehicles (AGVs) widely used in modern ports, unmanned warehouses and intelligent factories \cite{Choi2022AGV, Wang2022AGV,Zamora2021AGV}. Generally, AGV guidance methods can be divided into two types: fixed route and free route, based on different positioning technologies. In the traditional fixed route methods, a large number of magnetic markers or optical tapes prearranged in the planned path are commonly used \cite{Le-Anh2006AGV, Chen2018AGV}. These methods have the advantages of high reliability and robustness to disturbances, but are limited by their poor flexibility and high maintenance cost. The free route methods adopt more advanced positioning technologies based on radio, inertia, vision, laser and so on. They aim to provide guidance capability with flexible deployment, easy AGV route customization and low maintenance cost. However, these positioning technologies also face their own challenges of deterioration in accuracy and availability, e.g., error accumulation for the inertial navigation system, appalling weather for the laser-based system, weak illumination for the vision-based system, and 
blockages for the radio positioning system \cite{mohamed2019survey, Wang2021visible,An2022MEMA}.

\begin{figure}
	\centering
	\includegraphics[width=0.99\linewidth]{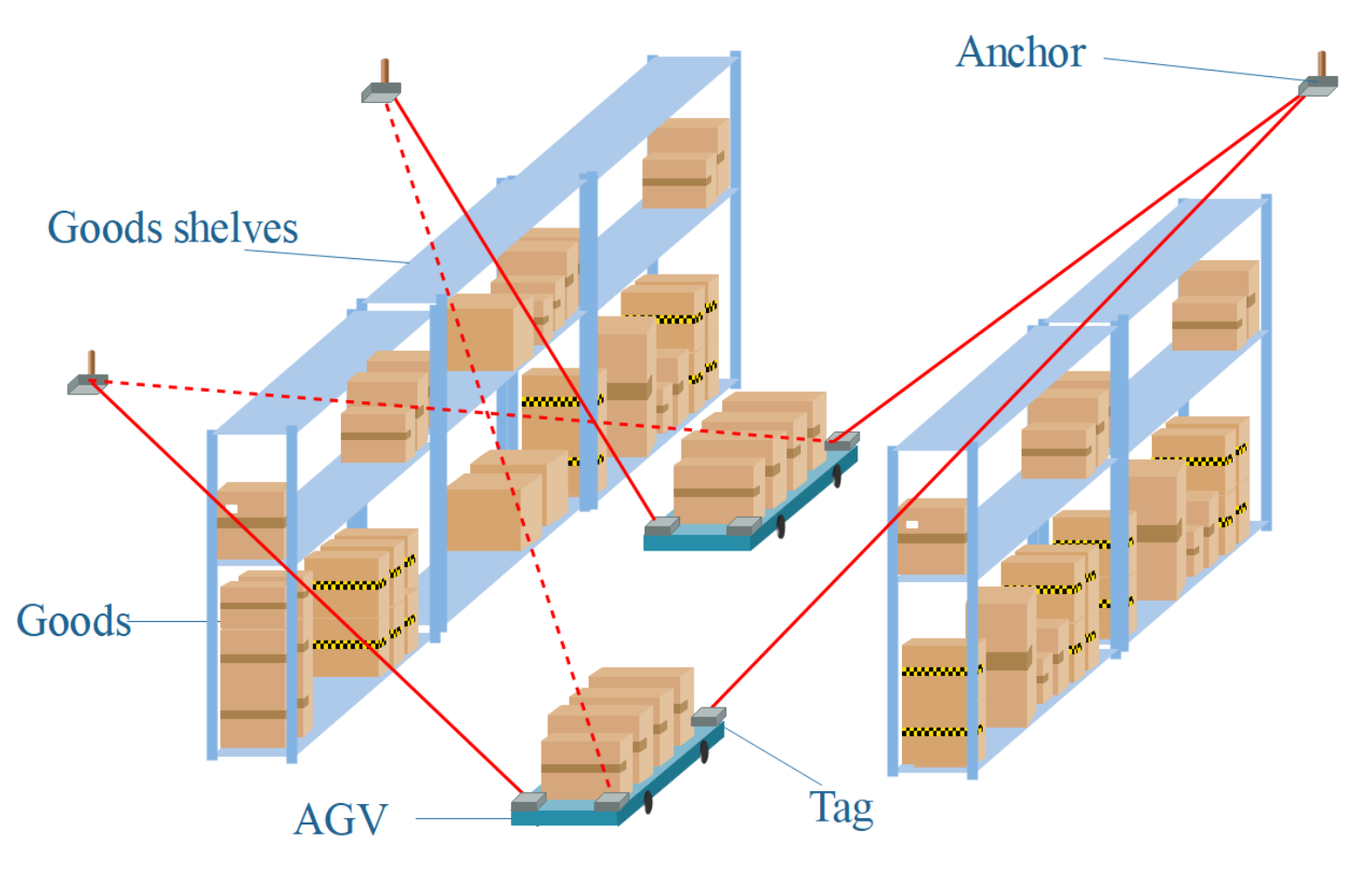}
	\vspace{-0.4cm}
	\caption{A typical AGV localization scenario in an unmanned warehouse. The blue and flat cuboid represents the AGV to be located, and the carton represents the loaded goods.
		The hanging gray cuboids and the ones mounted on the vehicles are the anchors and tags, respectively. The goods and shelves block the propagation of signals. The red solid line indicates the signal that can be received, while the red dotted line indicates the blocked signal.}
	\label{fig:scene_example}
	\vspace{-0.3cm}
\end{figure}

For free route methods, the range-based method that determine the position of the tag installed on the vehicle by trilateration or multilateration based on the available range measurements and the known anchor locations, is commonly utilized \cite{GuoNY2022NewClosedForm, Zhao2022Doppler, Zhao2022ClosedForm, Zhao2021OptimalTwoWay,Yang2022UWB}. The ranging system based on ultra-wideband (UWB) signals or lasers provide the range information between the tag and the anchors by measuring the time-of-flight (TOF) of the signals. They have a fine time resolution, and thus are widely used for vehicle localization. UWB signals are particularly well suited for both indoor and outdoor sub-meter level positioning applications \cite{Shen2010UWBfundamental, Zhao2021TOA,DJOSIC2022UWB}. However, due to the influences of the non-line-of-sight (NLOS) propagation and the limitations of the current physical layer technology, their ranging accuracy is limited. The system based on light detection and ranging (LiDAR) can also be used for TOF-based positioning method, on condition that the LiDAR scanner mounted on the vehicle is used as a tag and fluorescent reflectors are assigned as anchors \cite{Ronzoni2011AGVLidar,Shuai2018AGVrefle}. On the one hand, it will provide centimeter or even millimeter level ranging accuracy and contribute to higher positioning precision than the UWB system does. On the other hand, due to the periodic scanning mechanism, the computational complexity for LiDAR to obtain TOFs is much larger than that of the UWB system  \cite{Grollius2021lidar}, and its cost is much higher than the radio-based positioning system. Furthermore, it may face the problem of mutual interference caused by other LiDAR sensors \cite{Roriz2021AGVlidar}.

For the range-based positioning system, the number and quality of TOF measurements are decisive factors for accuracy and availability. Therefore, the obstacles from the environment and on the vehicle blocking the signals of radio or laser are regarded as the Achilles’ heel of TOF-based positioning systems. Fig. \ref{fig:scene_example} illustrates a typical AGV localization scenario in an unmanned warehouse. The vehicles transporting goods are navigated with the help of the positioning system. Goods on the shelves as well as loaded on the vehicles may block the propagation of signals. In such cases, a single tag obtains insufficient measurements for positioning and cannot guarantee the availability and accuracy \cite{Wymeersch2009CP, Nguyen2015LSCP}. 

To tackle the problem of insufficient measurements in such harsh environments, multi-antenna/receiver methods were proposed to improve positioning performance. Fan et al. \cite{fan2019dual-antenna} took the advantage of the spatial diversity between two antennas to determine the position and attitude of the vehicle and improve the availability and accuracy for GNSS receivers in urban environments. 
Multiple low-cost Global Positioning System (GPS) receivers were integrated by the extended Kalman filter at the pseudorange measurement level to provide accurate and robust position and attitude estimation \cite{shetty2015measurement}.
Ng and Gao \cite{Ng2017Multireceiver} presented a deeply coupled multi-receiver architecture to improve the reliability and robustness of position estimation by jointly tracking GPS signals received by multiple receivers.
By mounting multiple antennas with known positions on the AGV, a multi-antenna localization system based on TOA measurements was proposed to determine the position and attitude of the AGV in harsh environments \cite{An2020DMA}. 
However, they are all based on time-of-arrival (TOA) or pseudorange measurements, which contain the clock bias of the receiver, and all require strict inter-satellite or inter-anchor time synchronization. Thus, the above mentioned multi-antenna or multi-receiver methods are not applicable for TOF-based localization scenarios. 
 
Several methods based on TOF measurements were proposed to jointly estimate the position and attitude of a grid body, such as the rigid body localization (RBL)  \cite{Chepuri2014RBL_LS, Chen2015RBL_DAC, Jiang2018RBL_SDP,Jiang2019RBL_SDP2}. In RBL, a few sensors/tags are placed on the body and the measurements with respect to a few anchors are utilized to estimate the position vector and the rotation matrix between the body coordinate system (Frame b) and the global (computing/navigation) coordinate system (Frame g). However, the position vector and the rotation matrix are nonlinearly related to the measurements. Furthermore, the rotation matrix belongs to the special orthogonal (SO) group, implying non-convex constraints. Hence, the RBL problem is formulated as a difficult nonconvex optimization problem with nonconvex constraints. The existing solutions for RBL based on TOF measurements can be classified into two categorizes. The first category is the step-by-step approaches, e.g., the divide and conquer (DAC) method \cite{Chen2015RBL_DAC}. In this method, first, the global position of each tag is separately estimated by using only the TOF measurements of itself, and then the position and attitude of the rigid body are obtained according to the estimated tag positions. Although it has low complexity, this method does not have optimal positioning accuracy when the measurement noise is large or there are missing measurements, as will be shown in Section \ref{simulation}. The second category is the direct approaches, which estimate the position vector and rotation matrix directly \cite{Chepuri2014RBL_LS, Jiang2018RBL_SDP,Jiang2019RBL_SDP2}. Among them, the semi-definite relaxation (SDR) based method has the best robustness and accuracy \cite{Jiang2018RBL_SDP, Jiang2019RBL_SDP2}. They adopt SDR and auxiliary states to approximate the original problem to a semi-definite program (SDP) problem, and refine the solutions of the SDP problem with orthogonalization and corrections. The SDR based methods obtain good accuracy without requiring initial values. However, they have high computational complexity, since more parameters to be estimated are introduced in the approximated problem.
 
In this paper, we extend the classical RBL problem to the more challenging case with missing TOF measurements, corresponding to the multi-tag vehicle localization under harsh environments. 
We aim to find a solution with high availability, high accuracy and low complexity for this more general RBL problem. By modeling the system as a sensor network (SN) composed of two kinds of nodes, i.e., the anchors with known global coordinates and tags with known local coordinates, we use a structured Euclidean distance matrix (EDM) to represent both the exact and measured distances between nodes. On this basis, we propose an efficient RBL solution utilizing EDM completion approach, abbreviated as ERBL-EDMC.
Firstly, to reliably complete the original measured EDM with unspecified elements, we develop a new approach to determine the upper and lower bounds of the missing measurements by using the known local tag positions and the statistics of the available TOF measurements. Then, the global tag positions are obtained through coarse estimate based on the completed EDM and a refinement step assisted with inter-tag distances. Finally, the position and attitude of the vehicle are estimated via the iterative algorithm initialized by a closed-form result from the estimated global tag positions. 
Theoretical analysis and simulation results show that the proposed ERBL-EDMC method can effectively solve the RBL problem with missing measurements and obtain the optimal positioning results while maintaining low computational complexity compared with the state-of-the-art SDR based methods. To the best of the authors' knowledge, the new ERBL-EDMC is the first method to deal with the RBL problem with missing measurements.


The remainder of the article is organized as follows. Section \ref{problem} formulates the vehicle positioning problem. Section \ref{method_design} describes the design principle of the proposed ERBL-EDMC method. The proposed method is detailed in Section \ref{method}. Section \ref{performance} analyzes the availability, accuracy and computational complexity of the proposed method. Section \ref{simulation} presents the simulation results. And finally, the last section concludes the paper.

Main notations are summarized in Table \ref{table_notation}.

\begin{table}[!t]
	\caption{Notation List}
	\label{table_notation}
	\centering
	\begin{tabular}{l p{5cm}}
		\toprule
		lowercase $x$&  scalar\\
		bold lowercase $\boldsymbol{x}$ & vector\\
		bold uppercase $\bm{X}$ & matrix\\		
		$\hat{x}$, $\hat{\boldsymbol{x}}$ & estimated version of a variable\\
		$\tilde{x}$, $\tilde{\boldsymbol{x}}$ &  measured version of a variable\\	
		$\Vert \boldsymbol{x} \Vert$ & Euclidean norm of a vector\\	
		$\Vert \bm{X} \Vert$ & Frobenius norm of a matrix\\	
		$\bm{X}^T$, $\bm{X}^{-1}$ & matrix transpose and inverse, respectively\\	
		$\mathrm{det}(\cdot)$, $\mathrm{rank}(\cdot)$ & determinant and rank of a matrix, respectively \\
		$\operatorname{diag}(\bm{x}) $& diagonal matrix formed by the elelments of $\bm{x}$\\
    	$\operatorname{blkdiag}(\cdot) $& block diagonal matrix with the matrices inside along the diagonal\\
    	$\operatorname{null}(\bm{X}) $& an orthonormal basis of the null-space of $\bm{X}$\\
		$[\bm{X}]_{m,n}$ & element at the $m$-th row and  $n$-th column of a matrix\\
		$\otimes$ & Kronecker product\\     	
		$M$ & number of anchors\\
		$M_i$ & number of visible anchors of tag $i$\\
		$N$ & number of nodes in SN\\
	    $\bm{I}_K$ & $K\times K$ identity matrix\\
		$\bm{0}_{K}$ & $K$-element row vector with all-zero elements\\
		$\boldsymbol{1}_{K}$& $K$-element row vector with all-one elements\\
		$\bm{R}$&rotation matrix from Frame b to Frame g \\
		$\boldsymbol{l}_{i}$ & known local position vector of tag $i$\\
		$\boldsymbol{p}_{i}$ & unknown global position vector of tag $i$ \\
		$\boldsymbol{q}_{j}$ & known global position vector of anchor $j$\\
		$\boldsymbol{p}_{\mathrm{c}}$ &  unknown global position vector of the vehicle\\	
		$ \phi$, $\gamma$, $\psi$ &  pitch, roll and yaw angle of the body, respectively \\ 	
	    $\delta_{ij}$ & TOF measurement for tag $i$ from anchor $j$ \\
	    $d_{ij}$ & known distance from tag $i$ to tag  $j$ \\	
		$r_{ij}$ & known distance from anchor $i$ to anchor  $j$ \\		
		$\mathrm{card}(\mathcal{B})$ & cardinal number of elements in set  $\mathcal{B}$\\
		$\bm{F}$ &  Fisher information matrix (FIM)\\	
		\bottomrule
	\end{tabular}
\end{table}

\section{Problem Formulation} \label{problem}
We consider an AGV to be localized by a TOF-based positioning system in a harsh environment, in which, there are obstacles both in the environment and on the AGV itself, resulting in that some of the TOF measurements between the tags and anchors cannot be obtained. To obtain more available measurements, multiple tags with known positions relative to the local origin of Frame b are deliberately mounted on the vehicle.
This vehicle positioning problem is an RBL problem, in which the global coordinates of the origin of Frame b and the attitude of the vehicle need to be estimated by using the noisy and incomplete measurements between the tags and anchors.

We regard this system as an SN with a set of nodes $\mathcal{N}$ including the anchors and tags and $\operatorname{card}\left(\mathcal{N} \right)=N $. We denote the set of $M$ anchors by $\mathcal{A}$ and the known position of anchor $i$ by $ \bm{q}_{i},\ i \in\mathcal{A} $. The number of tags is $N-M$, and the tags are represented by set $\mathcal{T}$. The known position of tag $ i $ in Frame b  is $\bm{l}_i$, and the unknown position in Frame g is $\bm{p}_i$, $i\in\mathcal{T}$. According to the above definition, we have $\mathcal{A}\cup\mathcal{T}=\mathcal{N}$. Let $\mathcal{A}_{i}$ be the set of the visible anchors of tag $i$ with $\mathrm{card}(\mathcal{A}_{i})=M_{i}$, and $\mathcal{A}_{i}\subseteq	\mathcal{A}$.

The ranging measurements utilized here is TOF, which can be acquired with stringent time synchronization or multiple interactions between each tag-anchor pair. In this paper, we consider only the TOF measurements from the line-of-sight (LOS) or direct path and ignore the NLOS measurements since they can be identified and eliminated. The TOF measurements for tag $i$ from anchor $j$ is modeled as
\begin{equation}\label{eq:TOF}
\delta_{ij}=\Vert\bm{p}_{i}-\bm{q}_{j}\Vert+\varepsilon_{ij}\text{,} \quad i\in\mathcal{T}\text{,} \; j\in\mathcal{A},
\end{equation}
where $\delta_{ij} $ may not be available due to the obstacles and range limitation, the measurement noise $\varepsilon_{ij}$ is modeled as
independent and identically distributed Gaussian white noise with a variance of $\sigma^2$, i.e., $\varepsilon_{ij}\sim\mathrm{N}(0,\sigma^2)$.
 
The known distance from anchor $i$ to anchor $j$ is
  \begin{equation}\label{eq:d_anchor}
  	r_{ij}=\Vert\bm{q}_{i}-\bm{q}_{j}\Vert  	\text{,} \quad i\text{,} j\in\mathcal{A}\text{.}
  \end{equation}

According to the known positions of tags in Frame b, the inter-tag distances are known. The distance between tag $i$ and tag $j$ is modeled as 
\begin{equation}\label{eq:d_tag}
	d_{ij}=\Vert\bm{p}_{i}-\bm{p}_{j}\Vert =\Vert\bm{l}_{i}-\bm{l}_{j}\Vert 	\text{,} \quad i\text{,} j\in\mathcal{T}\text{,}
\end{equation}
which comes from the fact that the distance between two points remains unchanged through rigid transformation.

The relation between the tag position $\bm{p}_{i}$ in Frame g, the tag position $\bm{l}_{i}$ in Frame b and the position of the reference point on the vehicle (the origin of Frame b) $\bm{p}_{\mathrm{c}}$ in Frame g is \cite{Farrell2008Aided}
\begin{equation}\label{eq:pi}
	\bm{p}_{i}=\bm{p}_{\mathrm{c}}+\bm{R}\bm{l}_{i} \text{, } i \in\mathcal{T}
	\text{,}
\end{equation}
where $\bm{R}$ is the rotation matrix from Frame b to Frame g. 

We denote the attitude angle vector of the vehicle by $\bm{\theta}$, and $\bm{R}$ is then a function of $\bm{\theta}$ \cite{Farrell2008Aided}. For 3-dimensional (3D) problem, $\bm{\theta }=\left[\phi,\gamma ,\psi \right] $, denoting the roll, pitch and yaw angle, respectively. For 2-dimensional (2D) problem, the height, roll angle and pitch angle of the vehicle are considered as known constants and  $\theta =\psi$. 

To estimate the position $\bm{p}_{\mathrm{c}}$ and attitude $\bm{\theta} $ of the vehicle by using the available TOFs and the known tag positions in Frame b, we formulate the RBL problem as

\begin{align}
	\mathop {\min }\limits_{\bm{p}_{\mathrm{c}},{\bm{\theta }}} \sum\limits_{i\in \mathcal{T}} &\sum\limits_{j\in \mathcal{A}_i} v_{ij} {\left( {\left\| {\bm{p}_i - \bm{q}_{\mathcal{A}_{i}(j)}} \right\| - {\delta _{i\mathcal{A}_{i}(j)}}} \right)^2} \label{eq:original_problem} \\
	\operatorname{s.t.}\quad \bm {p}_i &=\bm{ p}_{\mathrm{c}} + \bm{R}\bm{l}_i\;\quad i \in \mathcal{T}, \tag{\ref{eq:original_problem}{a}} \label{eq:original_problema}\\ 
{\bm{R}} &= f\left( {\bm{\theta }} \right)\tag{\ref{eq:original_problem}{b}} \label{eq:original_problemb}\text{,}  
\end{align}
where the subscript ``$\mathcal{A}_{i}(j)$'' denotes the index of $j$-th element in set $\mathcal{A}_{i}$, $ v_{ij} $ represents the weight of the TOF measurement between tag $ i $ and the $ j $-th visible anchor of tag $i$ determined by its respective measurement noise, and $ f $ is a function of $\bm{\theta } $. 

Different from the RBL problem formulated in \cite{Chepuri2014RBL_LS, Chen2015RBL_DAC, Jiang2018RBL_SDP,Jiang2019RBL_SDP2}, the formulation in (\ref{eq:original_problem}) takes the case with missing measurements, due to the obstacles and range limitation in real-world applications, into consideration by employing ``$\mathcal{A}_{i}(j)$'' to index the anchor in the visible anchor set $\mathcal{A}_i$ of tag $i$. 

\section{Design Principle of ERBL-EDMC Method}\label{method_design}

\begin{figure}
	\centering
	\includegraphics[width=0.99\linewidth]{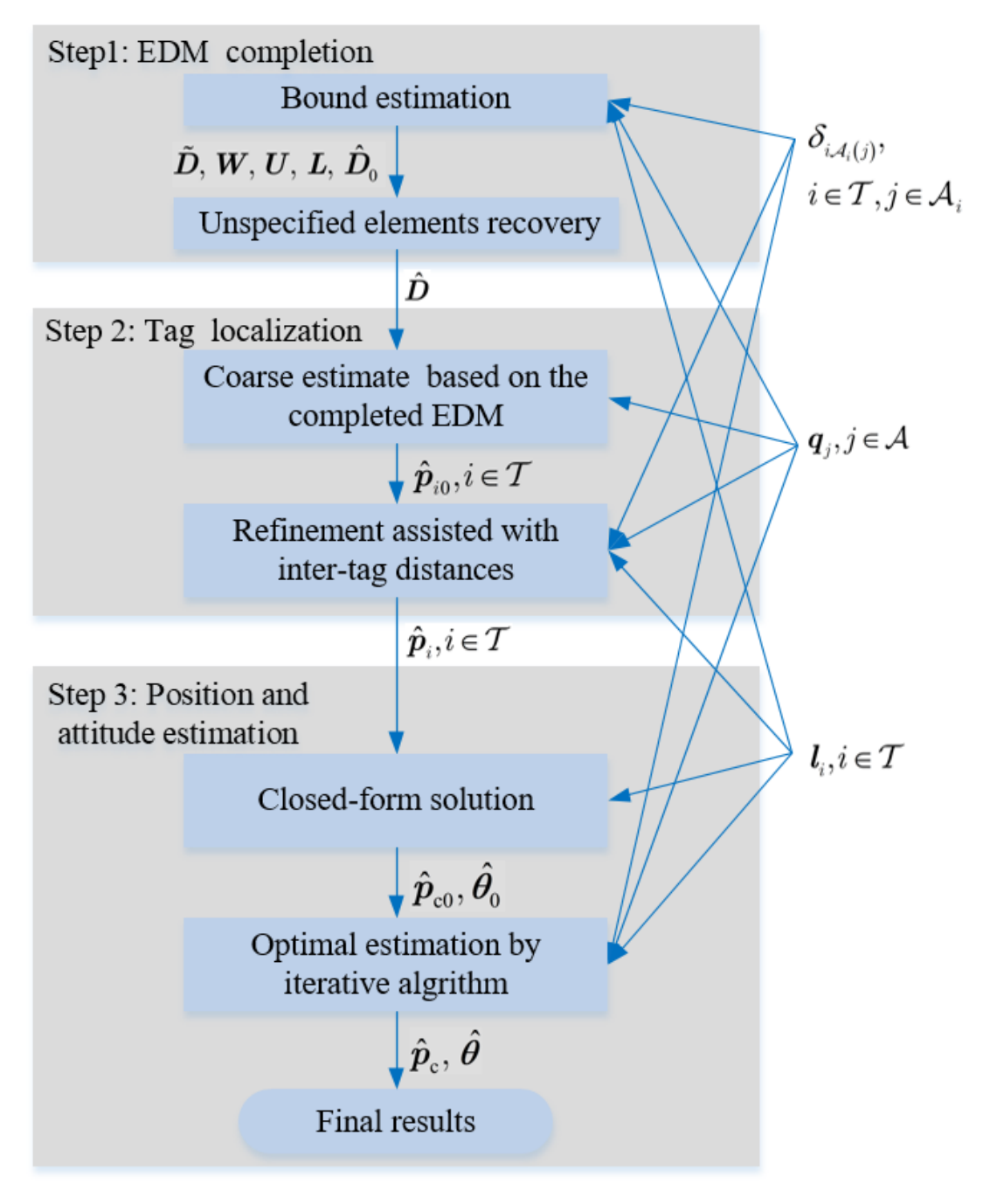}
	\vspace{-0.2cm}
	\caption{Flow chart of the ERBL-EDMC method. The proposed method is divided into three steps. The available TOFs $\delta_{i\mathcal{A}_i(j)}$, known global positions of anchors $\bm{q}_j$  and known local positions of tags $\bm{l}_i$ are applied to different steps to take the most advantage of existing information to obtain the desired positioning performance.}
	\label{fig:flow_diagram}
	\vspace{-0.3cm}
\end{figure}

The problem formulated in (\ref{eq:original_problem}) is a difficult nonconvex optimization problem with nonconvex constraints. Solving this problem based on the maximum likelihood estimate (MLE) achieves the asymptotic optimality, on the condition that it has an accurate initial guess to start an iterative implementation. Otherwise, the iteration may not converge or will be trapped at a local minimum. To obtain a good estimation for the position $\bm{p}_\mathrm{c}$ and attitude $\bm{\theta}$, initial values can be set based on the accurate tag locations as has been done in the DAC method \cite{Chen2015RBL_DAC}. However, different from the situation without missing TOFs in the DAC method \cite{Chen2015RBL_DAC}, the condition of insufficient TOFs in our RBL problem may degrade the accuracy or even disable the single tag position estimation. To overcome this difficulty, we intend to first recover the missing TOFs, then jointly estimate the tag positions with the assistance of inter-tag distances, and finally compute the vehicle's position and attitude.


First, we convert the recovery of the missing TOFs to a EDM completion problem. In order to ensure that the EDM completion can recover the unspecified elements reliably and efficiently, we develop a new approach for determining the upper and lower bounds of the missing TOFs, taking advantage of the known geometric relationships between anchors and between tags.

Then, the global positions of all tags are coarsely estimated by using the completed EDM from the previous step, and refined by using the measured TOFs as well as the known inter-tag distances. 

Finally, we obtain a closed-form solution of position vector and rotation matrix 
by using the global tag positions from the previous step. And then, substituting the closed-form solution into the optimization problem (\ref{eq:original_problem}) as the initial value, the final optimal estimates of the position and attitude of the vehicle are obtained iteratively.

To sum up, aiming to solve the problem (\ref{eq:original_problem}) effectively and accurately, we develop a new three-step RBL method, featuring a new EDM completion approach, and inter-tag distance assisted tag localization. We name it as ERBL-EDMC.

\section{Procedure of ERBL-EDMC Method}\label{method} 
As presented in Section \ref{method_design}, the ERBL-EDMC method consists of three main steps, which are shown in Fig. \ref{fig:flow_diagram}. The details of each step are described in the following subsections.

\subsection{Step 1: EDM Completion}
EDM is a matrix formed by the squared distances between some nodes in a $\eta$-dimensional Euclidean space. Since EDM implies the position relationship of the nodes and has a simple structure and useful properties, it is successfully applied in many applications such as sensor network localization, molecular conformation in chemistry and bioinformatics, and multidimensional scaling in statistics and machine learning \cite{morrison2003EDMapply,weinberger2006EDMapply,more1999EDMapply,glunt1993EDMapply }. For details on the definition, properties and applications of EDM, we refer the reader to \cite{Dokmanic2015EDM,krislock2012EDM,liberti2014EDM}. 

Due to some reasons such as communication range limitation, equipment failure, and propagation blockages, in many practical applications we cannot obtain all the pairwise distances. Therefore, the corresponding EDM is incomplete with some unspecified elements. The problem of determining the unspecified entries to recover an EDM is called EDM completion. This problem has attracted the attention in many areas of science and engineering and various algorithms have been proposed to solve it, mainly based on the low-rank property of EDM matrix \cite{moreira2018EDMC, fang2012EDMC, mishra2011EDMC}.

Solving the EDM completion problem is challenging due to the nonconvex rank constraint, the box constraint and possible local optima. A number of EDM completion method based on numerical optimization have been thoroughly studied \cite{alfakih1999SDP_EDM,pong2012SDP_EDM,ding2010SDP_EDM,more1997EDMC,fang2012EDMC,Zhou2018SQREDM}. The SDR based method \cite {alfakih1999SDP_EDM,pong2012SDP_EDM,ding2010SDP_EDM} which executes a global search exhibits great computational complexity. The EDM completion methods via the iterative algorithm require a good initialization to speed up the convergence and increase the chance of reaching the global minimum \cite{more1997EDMC,fang2012EDMC,Zhou2018SQREDM}. 
Among these, SQREDM \cite{Zhou2018SQREDM} is an efficient and low-complexity EDM completion approach. It adopts the conditional positive-semidefinite cone with rank cut that allows a fast computation of the projection onto this geometric object. This leads to the majorization-minimization algorithm that computes the transformed optimization problem elementwisely, and greatly reduces the computational complexity. 

In all these EDM completion methods\cite{alfakih1999SDP_EDM,pong2012SDP_EDM,ding2010SDP_EDM,more1997EDMC,fang2012EDMC,Zhou2018SQREDM}, the initial values or the bounds of unspecified elements are crucial for convergence and speed, and are estimated by the shortest path method based on undirected graph \cite{floyd1962shortest_path}. However, the bound estimates from the shortest path method may be far away from the real tag-anchor distances in harsh environment, especially for the case with obstacles on the vehicle itself. Hence, the EDM completion method will not converge or will converge to local optima, and degrades the EDM completion performance, as will be demonstrated later in the  simulations in Section  \ref{results_IEDM}. To this end, we develop a new bound estimation approach for EDM completion for the RBL problem in this step.

\subsubsection{EDM Completion Problem for RBL}
For the RBL issue, we construct a measured EDM $ \tilde{\bm{D}} $ by setting the square root of the elements to the measured or known distances modeled in (\ref{eq:TOF}), (\ref{eq:d_anchor}) and (\ref{eq:d_tag}). Let the weight matrix $\bm{W}$ represent the availability of the corresponding tag-anchor TOFs, and $[\bm{W}]_{m,n}$ is set to 1 if $ \delta _{(n-M)m} $ is available and 0 otherwise. Since $\bm{D}$ is symmetrical, we represent it by its upper triangular elements only, i.e., the elements with $ n \geqslant m $ as  
\begin{align}\label{eq:definition_D}
	&\sqrt{\tilde {[\bm{D}]}_{m,n}} =\\
	&\left\{
	\begin{matrix*}[l]
		\delta _{(n-M)m} &  [\bm{W}]_{m,n}=1,  m=1,\cdots,M,\\
	        	& n=M+1,\cdots,N,\\ 
	    r_{mn} & m=1,\cdots,M,\;m\leqslant n\leqslant M,\\
	 	d_{(m-M)(n-M)}&  m=M+1,\cdots,N,\; m\leqslant n\leqslant N,
	\end{matrix*}  \right.\nonumber
\end{align}
where $\tilde{[\bm{D}]}_{m,n}$ is the element at $m$-th row and $n$-th column of $\tilde{\bm{D}}$, and the elements corresponding to tag-anchor distances are the TOF measurements and the elements corresponding to the distances between anchors and between tags are the exact known distances. 
Note that with signal blockage and/or range limitation, some elements relating to the tag-anchor distances are unspecified and the corresponding $[\bm{W}]_{m,n}=0$, leading to an incomplete EDM.

\begin{figure}
	\centering
	\subfloat[Complete EDM]{
		\includegraphics[width=0.46\linewidth]{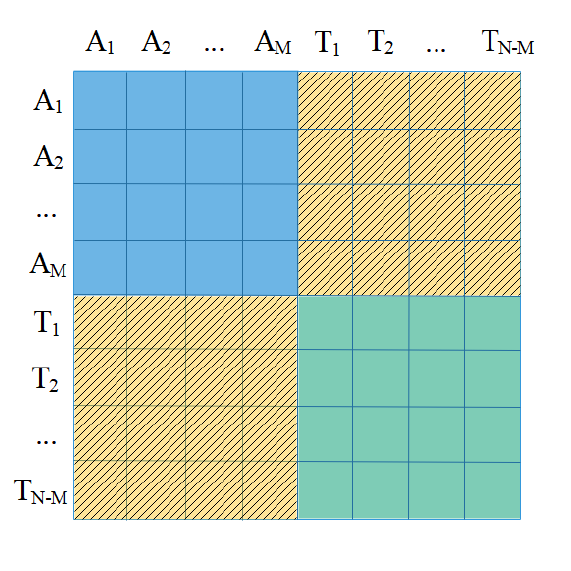}\label{fig:EDM1}}
	\hspace{0.02\linewidth}
	\subfloat[Incomplete EDM]{
		\includegraphics[width=0.46\linewidth]{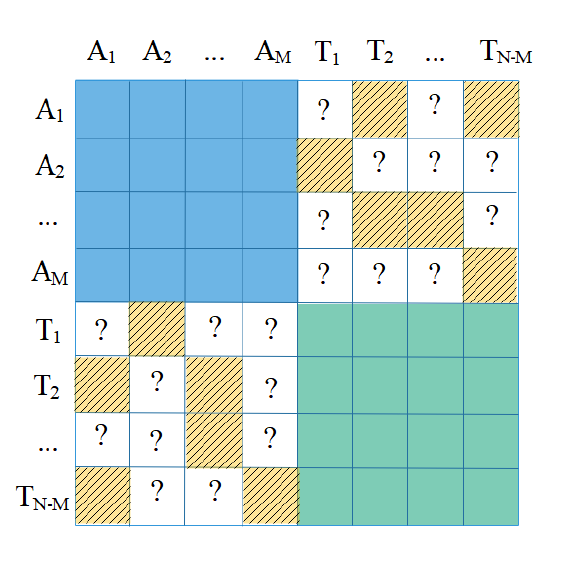}\label{fig:EDM2}}
	\vspace{0cm}
	\caption{EDM for RBL problem.
		(a) Complete EDM: all the elements are specified. (b) Incomplete EDM, in which some elements are unspecified due to the blockages  of the signal or laser propagation. Information included in EDM can be divided into three parts, the exact distances between anchors (denoted by ``A''), the exact distances between tags (denoted by ``T'') and the measured distances of tag-anchor pairs.}
	\label{fig:EDM}
	\vspace{-0.3cm}
\end{figure}

The illustrations of complete and incomplete EDMs for the RBL problem are shown in Fig. \ref{fig:EDM}. According to its definition, EDM has a block and symmetric structure comprised of three parts. The first part represented by the blue filled areas in Figs. \ref{fig:EDM1} and \ref{fig:EDM2} contains the known distances between anchors. The second part filled by green represents the known distances between tags. This is different from the conventional EDM applications, which employs noisy measurements between tags for the corresponding elements of the EDM. The last part illustrated by the shaded block relates to the noisy distance measurements between the tags and anchors. Furthermore, there are some ``?'' in this area in Fig. \ref{fig:EDM2}, representing the unspecified elements due to missing measurements. To obtain an accurate estimate of the tag positions, the original EDM $\tilde{\bm{D}} $ with noisy and unspecified elements must be completed and approximated to the real one. 

Due to the particularity that the inter-tag distances as well as the inter-anchor distances are known, we formulate the EDM completion problem for the RBL issue as
	\begin{align}\label{eq:EDM_problem}
		\mathop {\min }\limits_{\bm{D}} &\sum\limits_{m=1}^N\sum\limits_{n=1}^N {{[\bm{W}]}_{m,n}{{\left( {\sqrt {{{[\bm{D}]}_{m,n}}}  - \sqrt{\tilde{[\bm{D}]}_{m,n}}} \right)}^2}}   \\
		\nonumber \operatorname{s.t.} \quad &  \operatorname{rank}({\bm{JDJ}}) \leqslant \eta,\ {\bm{D}} \in \mathtt{D}^N,\bm{L} \preceq \bm{D}\preceq \bm{U},
	\end{align}		 
in which,
$\bm{W}$ is defined as in (\ref{eq:definition_D}), $ \mathtt{D}^N $ denotes the set of EDMs of size $ N\times N $, the rank constraint $ \operatorname{rank}({\bm{JDJ}})\leqslant \eta $ comes from the rank property of EDM \cite{Dokmanic2015EDM}, ${\bm{J}}  \triangleq {\bm{I}_N} - \frac{1}{N}{\bm{1}^T_N}{{\bm{1}_N}}$, $ \bm{I}_N $ is the $ N\times N $ identity matrix and $\bm{1}_N$ denotes the $ N$-element row vector of all-one elements,  and $ \bm{L} $ and $\bm{U}$ are the box constraint matrices formed by the squared lower and upper bounds of the distances, respectively, which help to describe the EDM completion problem more accurately and reduce the search space. 



We formulate the EDM completion problem as (\ref{eq:EDM_problem}) by taking only the tag-anchor measurements into the optimization cost function. Furthermore, a new bound estimation is developed to determine the lower and upper bounds of the missing tag-anchor TOFs.



\subsubsection{Bound Estimation}

As shown in Fig. \ref{fig:flow_diagram}, we develop a new bound estimation approach to determine the upper and lower bounds of the missing measurements 
based on the geometric relationship between different tags with the same visible anchors.

Specifically, for the EDM elements representing the known inter-anchor and inter-tag distances, the lower and upper bounds are the same and are set to their true values, i.e., $r_{ij},\;i,j \in \mathcal{A}$ and $d_{ij}, \;i,j \in \mathcal{T}$. For the elements representing tag-anchor distances with available TOF measurements, the lower and upper bounds are set as the $\pm3 \sigma$ of the measured value, i.e, $\delta _{ij} - 3\sigma$ and $\delta _{ij} + 3\sigma$.

Most importantly, for the missing TOFs, the lower and upper bounds are set according to the relationship between different tags and their commonly visible anchors. Based on the triangle inequality, we have
\begin{align}
\label{eq:triangle}
\left\|\bm{p}_k-\bm{q}_j\right\|-d_{ki}\leqslant\left\|\bm{p}_i-\bm{q}_j\right\|\leqslant \left\|\bm{p}_k-\bm{q}_j\right\|+d_{ki},\nonumber\\
j\in \mathcal{A}, \;i,k\in \mathcal{T}, \;k\ne i.
\end{align}

Taking the $\pm3 \sigma$ of the measurement noise into consideration, we can replace the true tag-anchor distances in \eqref{eq:triangle} by the TOF measurement $\delta_{ik}\pm3 \sigma$. Note that $\bm{L}$ and $\bm{U}$ are symmetrical. Then, utilizing the measured and known distances in (\ref{eq:TOF}), (\ref{eq:d_anchor}) and (\ref{eq:d_tag}), the lower and upper bounds can be represented by their upper triangular elements, 
as

\begin{align}\label{eq:L}
 	 	&\sqrt{{[\bm{L}]}_{m,n}} = \\
 	 	&\left\{ \begin{matrix*}[l]
 	  		0& m = n,\\
 	{r_{mn}}	&  m=1,\cdots,M,\;m < n\leqslant M, \\
 			{d_{(m-M)(n-M)}}&  m=M+1,\cdots,N, \;m < n\leqslant N, \\
 		\delta _{(n-M)m}-3\sigma 	& [\bm{W}]_{m,n}=1,\;m=1,\cdots,M,\\
		& n=M+1,\cdots,N ,		 \\ 
		\operatorname{min}\hat{\delta} _{(n-M)m}-3\sigma& 
 		  [\bm{W}]_{m,n}=0, \;m=1,\cdots,M,\\
		&n=M+1,\cdots,N, \\ 
\end{matrix*} \right. \hfill \nonumber
 \end{align}

\begin{align}\label{eq:U}
 	 	&\sqrt{{[\bm{U}]}_{m,n}} = \\
 	 	&\left\{ \begin{matrix*}[l]
 	  		0& m = n,\\
 		{r_{mn}}& m=1,\cdots,M,\;m < n\leqslant M,\\
 	 			{d_{(m-M)(n-M)}}& m=M+1,\cdots,N, \; m < n\leqslant N,\\
 			\delta _{(n-M)m}+3\sigma & [\bm{W}]_{m,n}=1,\;m=1,\cdots,M,\\ &n=M+1,\cdots,N,	\\ 
		\operatorname{max}\hat{\delta} _{(n-M)m}+3\sigma&   [\bm{W}]_{m,n}=0,\; m=1,\cdots,M,\\
		&n=M+1,\cdots,N,
\end{matrix*}  \right. \hfill \nonumber
 \end{align} 
where the minimum and maximum value for the missing measurement estimated based on (\ref{eq:triangle}) are 
\begin{align}
\nonumber\operatorname{min}&\hat{\delta}_{(n-M)m}=\\
&\max \left\{ {\delta _{km}} - {d_{(k-M)(n-M)}}\left| k=M+1,\cdots,N,\right.\right.\nonumber\\
&\left.\left.k \ne n, [\bm{W}]_{k,n} = 1 \right. \right\},\nonumber
\end{align}
and
\begin{align}
\operatorname{max}&\hat{\delta}_{(n-M)m}=\nonumber\\
&\min\left\{ {\delta _{km}} + {d_{(k-M)(n-M)}}\left| k=M+1,\cdots,N,\right.\right.\nonumber\\
&\left.\left.k \ne n, [\bm{W}]_{k,n} = 1 \right. \right\}.\nonumber
\end{align}

Utilizing  $\bm{L}$ and $\bm{U}$, we set the initial EDM as
\begin{align}\label{eq:initD}
  \hat{\bm{D}}_0=\frac{\bm{L}+\bm{U}}{2}.
\end{align}
 
Substituting $ \hat{\bm{D}}_0 $, $  \bm{L}$, $ \bm{U}$ and $\bm{W}$ into problem (\ref{eq:EDM_problem}), we can obtain the completed EDM $ \hat{\bm{D}}$ by utilizing the conventional EDM completion method such as \cite{more1997EDMC,fang2012EDMC,Zhou2018SQREDM}. Due to its efficiency and low complexity, the SQREDM method \cite{Zhou2018SQREDM} is selected.  
 
\subsection{Step 2: Tag Localization}
In this step, the estimation of the global tag positions are initialized by the coarse estimation based on the completed EDM $ \hat{\bm{D}}$ from Step 1 and then refined iteratively based on the TOF measurements with the assistance of inter-tag distances. 

With the completed EDM $ \hat{\bm{D}}$, the distance information of each tag-anchor pair is available.
Different from the conventional DAC method, which
only uses measurements of a single tag and thus may have degraded localization accuracy under a poor geometry \cite{Chen2015RBL_DAC}, we jointly estimate the tag positions based on the multi-tag TOFs assisted by the inter-tag distance to improve the geometry of the overall visible anchors and provide more information for the estimation. 
 
We formulate the inter-tag distance assisted tag localization problem as
\begin{equation}\label{eq:problem_CL}
\begin{aligned}
 \underset{\bm{p}_i\;i\in\mathcal{T}}{\operatorname{argmin}}&\left\lbrack \sum\limits_{i\in \mathcal{T} }\sum\limits_{j \in \mathcal{A}_i}\alpha_{ij}\left( \delta_{i\mathcal{A}_i(j)} - \Vert\bm{p}_{i}-\bm{q}_{\mathcal{A}_i(j)}\Vert \right)^{2}\right.\\
 &\left.+\sum\limits_{i\in \mathcal{T} } \sum\limits_{k\in \mathcal{T},k>i }\beta_{ik}\left( d_{ik} - \Vert \bm{p}_{i} - \bm{p}_{k} \Vert \right)^{2} \right\rbrack,
\end{aligned}
\end{equation}
where $\alpha_{ij} $ and $\beta_{ik}$ are the weights for the TOF measurements and the distances between tags, respectively. According to the assumptions for TOF measurement noises, we set $\alpha_{ij} = \sigma^{- 2} $ and $\beta_{ik} = \lambda \sigma^{- 2} $, where $\lambda $ is a factor greater than 1, representing that the inter-tag distances are set with greater weights, since they have higher accuracy.

Then, we present an iterative method to obtain the global tag position estimates. We write the unknown tag positions in the collective form as
$$\bm{y}= \left[\begin{matrix} \bm{p}_1^T & \bm{p}_2^T & \cdots &\bm{p}_{N-M}^T \end{matrix}\right]^T.
$$

We linearize (\ref{eq:TOF}) and (\ref{eq:d_tag}) by Taylor expansion and keep the first order terms, substitute them into (\ref{eq:problem_CL}) and have
 \begin{equation}
\label{eq:problem_CL_linear}
\underset{\Delta\bm{y}}{\operatorname{argmin}}
\left( {\Delta\bm{z} - \bm{H}\Delta\bm{y}} \right)^{T}\bm{Q}\left( {\Delta\bm{z} - \bm{H}\Delta\bm{y}} \right),
\end{equation}
where the weighting matrix $\bm{Q} $ is
$$\bm{Q} = \operatorname{blkdiag}(\sigma^{- 2}\bm{I}_{{\delta}}, {\lambda\sigma^{- 2}\bm{I}_{d}}),$$
in which the TOF related  $\bm{I}_{{\delta}} $  is $\sum\limits_{i\in \mathcal{T}}M_{i} \times \sum\limits_{i \in \mathcal{T}}M_{i} $ identity matrix, the inter-tag distance related  $\bm{I}_{d} $ is $\sum\limits_{i\in \mathcal{T}}(N-M-i)\times \sum\limits_{i\in \mathcal{T}}(N-M-i) $ identity matrix,
 the increment $$\Delta\bm{y} = \bm{y} - \hat{\bm{y}}, $$ 
 the residual  $$\Delta\bm{z} = \left[ \begin{matrix} \Delta \bm{z}_{\delta}^T & \Delta \bm{z}_{d}^T \end{matrix}\right]^T,$$
and the design matrix
$$\bm{H} = \left[ \begin{matrix} \bm{H}_{\delta}^T & \bm{H}_{d}^T \end{matrix} \right]^T,$$ in which
\begin{equation}
\begin{aligned}
\Delta \bm{z}_{\delta}&= \left[\begin{matrix} \Delta \bm{z}_{\delta,1}^T & \Delta \bm{z}_{\delta,2}^T & \cdots &\Delta \bm{z}_{\delta,{N-M}}^T \end{matrix}\right]^T,\\
\Delta \bm{z}_{d}&= \left[\begin{matrix} \Delta \bm{z}_{d,1}^T & \Delta \bm{z}_{d,2}^T & \cdots &\Delta \bm{z}_{d,{N-M}}^T \end{matrix}\right]^T,\\ 
\bm{H}_{{\delta}} 
&= \operatorname{blkdiag}\left(
	\bm{H}_{\delta,1}, \bm{H}_{\delta,2},\cdots,\bm{H}_{\delta,N-M} 
	 \right),\\
\bm{H}_{d}&= \left[\begin{matrix}
	\bm{H}_{d,1}^T& \bm{H}_{d,2}^T&\cdots&\bm{H}_{d,N-M}^T\end{matrix}
	 \right]^T,	 \nonumber
\end{aligned}
\end{equation}
the parameters related to the TOF measurements and the distances between tags are denoted with subscripts ``$\delta $'' and ``$d$'', respectively, for tag $i,\;i\in\mathcal{T}$, the corresponding residual and design matrix are
 \begin{equation}
 \begin{aligned}
\Delta \bm{z}&_{\delta,i}= \\
 &\left[ {\delta}_{i\mathcal{A}_i(1)} - {\Vert{\hat{\bm{p}}_{i} - {\bm{q}}_{\mathcal{A}_i(1)}}\Vert}\quad {\delta}_{i\mathcal{A}_i(2)} - {\Vert{\hat{\bm{p}}_{i} - {\bm{q}}_{\mathcal{A}_i(2)}}\Vert} \quad \cdots \right.\\ &\left.\quad{\delta}_{i\mathcal{A}_i(M_i)} - {\Vert{\hat{\bm{p}}_{i} - {\bm{q}}_{\mathcal{A}_i(M_i)}}\Vert}\right]^T,  \nonumber
 \end{aligned}
 \end{equation}
  \begin{equation}
 \begin{aligned}
\Delta \bm{z}&_{d,i}= \\
 &\left[ d_{i(i+1)} - {\Vert{\hat{\bm{p}}_{i} - \hat{\bm{p}}_{i+1}}\Vert}\quad {d}_{i(i+2)} - {\Vert{\hat{\bm{p}}_{i} -\hat{\bm{p}}_{i+2}}\Vert} \quad \cdots \right.\\ &\left.\quad{d}_{i(N-M)} - {\Vert{\hat{\bm{p}}_{i}  -\hat{\bm{p}}_{N-M}}\Vert}\right]^T,  \nonumber
 \end{aligned}
 \end{equation}
\begin{equation}
   \bm{H}_{\delta,i} = \left[ \begin{matrix} \frac{\hat{\bm{p}}_{i} - \bm{q}_{\mathcal{A}_i(1)}}{\Vert{\hat{\bm{p}}_{i} - {\bm{q}}_{\mathcal{A}_i(1)}}\Vert} & \frac{\hat{\bm{p}}_{i} - \bm{q}_{\mathcal{A}_i(2)}}{\Vert{\hat{\bm{p}}_{i} - {\bm{q}}_{\mathcal{A}_i(2)}}\Vert} &\cdots&\frac{\hat{\bm{p}}_{i} - \bm{q}_{\mathcal{A}_i(M_i)}}{\Vert{\hat{\bm{p}}_{i} - {\bm{q}}_{\mathcal{A}_i(M_i)}}\Vert} \end{matrix} \right]^T,\nonumber
   \end{equation}
 and 
 \begin{equation}
 \begin{aligned}
&\bm{H}_{d,i}=\\
&\left[ {\begin{matrix}
    {{{\bm{0}}_{\eta\cdot(i-1)}}}
&-\bm{u}_{i(i+1)}&
  \bm{u}_{i(i+1)}
& {\bm{0}}_{\eta}& {\bm{0}}_{ \eta}&
  \cdots\\
  {{{\bm{0}}_{ \eta\cdot(i-1)}}}
&-\bm{u}_{i(i+2)}&
 {\bm{0}}_{\eta} 
& \bm{u}_{i(i+2)}& {\bm{0}}_{\eta}&
\cdots\\
 \vdots
&\vdots&
  \vdots
& \ddots& \ddots&
 \vdots\\
 {{{\bm{0}}_{ \eta\cdot(i-1)}}}
&-\bm{u}_{i(N-M)}&
   {\bm{0}}_{\eta} 
& \cdots& {\bm{0}}_{ \eta}&
  \bm{u}_{i(N-M)} 
\end{matrix}} \right],\nonumber
\end{aligned}
\end{equation}
in which, $\bm{H}_{d,i}$ is a $ (N-M-i)\times \eta\cdot(N-M)$ matrix, $\bm{0}_\kappa$ is a $\kappa$-element row vector with all-zero elements and ${\bm{u}}_{ik}=\frac{(\hat{\bm{p}}_{i} - \hat{\bm{p}}_{k})^T}{d_{ik}},\;k\in\mathcal{T},k>i$.
 
We estimate $\Delta\bm{y} $ in a weighted least squares sense as
 \begin{equation}\label{eq:CL_deltay}
\Delta\hat{\bm{y}} =  \left( {\bm{H}^{T}\bm{Q}\bm{H}} \right)^{- 1}\bm{H}^{T}\bm{Q}\Delta\bm{z}.
\end{equation}

The accurate tag position estimation $\hat{\bm{y}}$ can then be obtained iteratively given a fine initialization. Note that the position of each tag can be estimated separately by utilizing the completed EDM  $\hat{\bm{D}}$. Hence, we coarsely estimate the tag positions $\hat{\bm{p}}_{i0}, i \in \mathcal{T}$ by the classical localization methods based on ranging measurements \cite{foy1976positioning, chan1994positioning, manolakis1996positioning} and set them as the initial value for the iterative algorithm to obtain a refined estimation.

\subsection{Step 3: Position and Attitude Estimation}\label{method_RBL}
We aim to obtain the position vector $\bm{p}_\mathrm{c}$ and the attitude $\bm{\theta}$ of the vehicle. Taking advantage of the global tag coordinates obtained in Step 2 and the known local tag coordinates, the position and rotation matrix estimation problem can be formulated as \cite{Chen2015RBL_DAC} 
\begin{align}\label{eq:Procrustes}
 &\mathop {\min }\limits_{\bm{p}_\mathrm{c},\bm{R}} \sum\limits_{i \in \mathcal{T}} {{{\left( {\hat{\bm{p}}_i -\bm {p}_\mathrm{c} - \bm{R}{{\bm{l}}_i}} \right)^T\Xi_i \left( {\hat{\bm{p}}_i -\bm {p}_\mathrm{c} - \bm{R}{{\bm{l}}_i}} \right)}}}, \\
	&\operatorname{s.t.} \quad   {\bm{R}^T}\bm{R} =\bm{I},\;\operatorname{det}  \left( \bm{R} \right) = 1, \label{eq:SOcstr}
\end{align}
where the weighting matrix $\Xi_i$ is the inverse of covariance of $\hat{\bm{p}}_i$. 

Theoretically, 
if $\Xi_i$  is the Fisher information matrix (FIM), the optimization (\ref{eq:Procrustes}) yields the optimal position and attitude estimates \cite{abel1989DAC,Chen2015RBL_DAC}. However, since $\bm{p}_\mathrm{c}$ and $\bm{R}$ are unknown, the FIM is not available at this stage. Therefore, to solve (\ref{eq:Procrustes}), we set $\Xi_i$ as a unity matrix, which leads to a loss in accuracy, to get a closed-form solution, 
and then obtain the optimal solution iteratively.

The closed-form solution for the rotation matrix $\bm{R}$, denoted by $\hat{\bm{R}}_0$, is obtained by \cite{Eggert1997transformation}
\begin{equation}\label{eq:R}
    \hat{\bm{R}}_0 =\bm{V}_1\operatorname{diag} \left(\left[\begin{matrix}
  \bm{1}_{\eta-1}&\operatorname{det}\left( \bm{V}_1\bm{V}_2^T\right) \end{matrix}\right] \right)
 \bm{V}_2^T,
\end{equation}
where $\bm{V}_1$ and $\bm{V}_2$ come from the singular value decomposition (SVD) 
\begin{equation}
    \bm{V}_2\bm{\Sigma} \bm{V}_1^T = \operatorname{SVD}\left( \sum\limits_{i \in \mathcal{T}}(\bm{l}_i-\bar{\bm{l}})(\hat{\bm{p}}_i-\bar{\bm{p}})^T \right).
\end{equation}
The closed-form solution for the position vector $\bm{p}_\mathrm{c}$, denoted by $\hat{\bm{p}}_{\mathrm{c}0}$, is then calculated by \cite{Chen2015RBL_DAC}

\begin{equation}\label{eq:p_R}
\hat{\bm{p}}_{\mathrm{c}0} = \bar  {\bm{p}} - \hat{\bm{R}}_0\bar {\bm{l}},
\end{equation}
where $\bar {\bm{p}} = \frac{1}{N-M}\sum\limits_{i \in \mathcal{T}} {{\hat{\bm{p}}_i}} $ and $\bar {\bm{l}} = \frac{1}{N-M}\sum\limits_{i \in \mathcal{T}} {{\bm{l}_i}} $.

Due to the errors in tag position estimates and the approximation of $\Xi_i$ in solving (\ref{eq:Procrustes}) as analyzed above, the results of (\ref{eq:R}) and (\ref{eq:p_R}) can not be directly used in high-precision applications. We set the initial values $\hat{\bm{p}}_{\mathrm{c}0}$ and $\hat{\bm{\theta}}_0$ from the closed-form solutions (\ref{eq:p_R}) and (\ref{eq:R}), and employ an iterative algorithm to solve the original problem (\ref{eq:original_problem}) to estimate the position $\bm{p}_\mathrm{c}$ and the attitude angle $\bm{\theta}$. 

We write the unknown position $\bm{p}_\mathrm{c}$ and attitude angle $\bm{\theta}$ of the vehicle in the collective form as
$$\bm{x}= \left[\begin{matrix} \bm{p}_\mathrm{c}^T & \bm{\theta}^T  \end{matrix}\right]^T.
$$

Utilizing the constraints (\ref{eq:original_problema}) and (\ref{eq:original_problemb}), we linearize (\ref{eq:TOF}) with respect to $\bm{x}$ by Taylor expansion, substitute them into the cost function of problem (\ref{eq:original_problem})  and have
\begin{equation}
	\label{eq:original_problem_linear}
	\underset{\Delta\bm{x}}{\operatorname{argmin}}
	\left( {\Delta\bm{\zeta} - \bm{G}\Delta\bm{x}} \right)^{T}\bm{P}\left( {\Delta\bm{\zeta} - \bm{G}\Delta\bm{x}} \right),
\end{equation}
where the weighting matrix $\bm{P} $ is
$$	\bm{P} = \sigma^{- 2}\bm{I}_{{\delta}},$$
the increment $$\Delta\bm{x} = \bm{x} - \hat{\bm{x}}, $$ 
the residual  $$	\Delta \bm{\zeta}= \left[\begin{matrix} \Delta \bm{\zeta}_{1}^T & \Delta \bm{\zeta}_{2}^T & \cdots &\Delta \bm{\zeta}_{{N-M}}^T \end{matrix}\right]^T,$$ 
and the design matrix
$$	\bm{G}
= \left[ \begin{matrix}
\bm{G}_{1}^T&\bm{G}_{2}^T&\cdots&\bm{G}_{N-M} ^T
\end{matrix}\right]^T,$$
in which, for tag $i,\;i\in\mathcal{T}$, the corresponding residual is
 \begin{equation}
 \begin{aligned}
\Delta \bm{\zeta}_{i}= \left[ \begin{matrix}
{\delta}_{i\mathcal{A}_i(1)} - {\Vert{\hat{\bm{p}}_\mathrm{c}+\hat{\bm{R}}\bm{l}_i - {\bm{q}}_{\mathcal{A}_i(1)}}\Vert}\\
 {\delta}_{i\mathcal{A}_i(2)} - {\Vert{\hat{\bm{p}}_\mathrm{c}+\hat{\bm{R}}\bm{l}_i-{\bm{q}} _{\mathcal{A}_i(2)}}\Vert} \\
 \vdots \\ {\delta}_{i\mathcal{A}_i(M_i)} - {\Vert{\hat{\bm{p}}_\mathrm{c}+\hat{\bm{R}}\bm{l}_i - {\bm{q}}_{\mathcal{A}_i(M_i)}}\Vert}
\end{matrix}\right],  \nonumber
 \end{aligned}
 \end{equation}
the sub-matrix $\bm{G}_i$ of the design matrix is an $M_i\times (\eta+\nu) $ matrix as
$$\bm{G}_i = \left[ \begin{matrix} {\frac{{\partial {g_{i1}}({\bm{x}})}}{{\partial {\bm{x}}}}}& {\frac{{\partial {g_{i2}}({\bm{x}})}}{{\partial {\bm{x}}}}}&\cdots&{\frac{{\partial {g_{iM_i}}({\bm{x}})}}{{\partial {\bm{x}}}}}\end{matrix}\right]^T,$$ 
 where $\nu$ is the number of attitude angles to be estimated (1 for 2D case and 3 for 3D case), and $g_{ij} (\bm{x})={\left\| {\bm{p}_\mathrm{c}+\bm{R}\bm{l}_i - \bm{q}_{\mathcal{A}_{i}(j)}} \right\| - {\delta _{i\mathcal{A}_{i}(j)}}}, \;j\in \mathcal{A}_i$.
 
We estimate $\Delta\bm{x} $ in a weighted least squares sense as
\begin{equation}\label{eq:deltax}
	\Delta\hat{\bm{x}} =  \left( {\bm{G}^{T}\bm{P}\bm{G}} \right)^{- 1}\bm{G}^{T}\bm{P}\Delta\bm{\zeta}.
\end{equation}

And then, the optimal solutions for the position and attitude of the vehicle are obtained by iteratively updating the initial values $\hat{\bm{p}}_{\mathrm{c}0}$ and $\hat{\bm{\theta}}_0$. We adopt the Gauss-Newton method to iteratively compute the optimal solutions. The iterative algorithm stops if the norm of the increment of the current iteration is lower than a preset threshold $\xi$ , i.e., $\Vert\Delta\hat{\bm{x}}_q  \vert<\xi $ for the $q$-th iteration\cite{Nocedal1999Numerical,Kay1993SS}. Otherwise, the algorithm enters the $(q+1)$-th  iteration, updates $\hat{\bm{x}}_{q+1}=\hat{\bm{x}}_q+\Delta\hat{\bm{x}}_q$, and continues to calculate $\Delta\hat{\bm{x}}_{q+1}$ by \eqref{eq:deltax}. Furthermore, for the case that the iterative algorithm does not converge, a maximum number of iterations  $q_{\operatorname{max}}$ is set. When the iteration count $q$ is greater than  $q_{\operatorname{max}}$, the algorithm stops and indicates a failure in obtaining the location solution. The parameter $\xi$  is set to $10^{-8}$,   and $q_{\operatorname{max}}$  is set to 20 in our simulations.
 
\section{Performance Analysis}\label{performance}
In this section, we evaluate the performance of the proposed ERBL-EDMC method by theoretically analyzing the availability, accuracy and computational complexity.
\subsection{Availability}
Availability indicates whether the positioning service is available, given the needed accuracy.
Taking the 2D case for example, the parameters to be estimated are the 2D coordinates and the yaw angle, i.e., three unknowns are to be determined. Therefore, the corresponding requirement is that the total number of TOFs for all the tags must be no less than three. In addition, to estimate the yaw angle, there should be no less than two tags, that each can observe at least one anchor. Hence, the availability condition is
 \begin{equation}
 \label{eq:available}
 	\sum\limits_{i\in\mathcal{T}} M_i\geqslant 3 \quad \text{and} \quad \mathrm{card}(\mathcal{C})\geqslant 2 
 \end{equation} 
where $ \mathcal{C} = \left\lbrace M_i>0,\; i\in \mathcal{T} \right\rbrace$ is the set of tags that have TOF measurements.

Unlike the single tag positioning \cite{Guvenc2009TOFpositioning, Liu2007TOFpositioning}, which requires sufficient measurements for each tag, the proposed method only requires a sufficient total number of measurements for all the tags. In addition, for the conventional DAC method, at least 3 measurements are required for each tag to ensure a unique positioning solution. Therefore, compared with the single tag positioning and the DAC methods, our method has more relaxed requirements for measurements, and thus has better availability.
\subsection{Accuracy}
For the new ERBL-EDMC method, we aim to estimate the position and attitude of the AGV efficiently and accurately. The results of the tag positions determine the accuracy of the closed-form solution in Step 3 and thus affect the
convergence speed and estimation error of the iterative algorithm. Hence, we first analyze the accuracy of tag localization and then derive the CRLB of position and attitude estimation in this subsection.  
\subsubsection{Accuracy of Tag Localization}
In Step 2 of the ERBL-EDMC method, the global positions of the tags are estimated based on the TOF measurements and the known inter-tag distances.

Following the derivation in \cite{Huang2016CP}, we have the covariance matrix of the positioning error for the tag position as
\begin{equation} \label{eq:CRLB_CL}
	\operatorname{cov}(\Delta\hat{\bm{y}})=\sigma^2\left( \bm{H}_{\delta}^T\bm{H}_{\delta}+\frac{1}{\lambda}\bm{H}_d^T\bm{H}_d\right) ^{-1},
\end{equation}
where $\bm{H}_{\delta}$, $\bm{H}_d$, $\sigma$ and $\lambda$ are the same as in (\ref{eq:problem_CL_linear}).

According to (\ref{eq:CRLB_CL}), the estimate accuracy for each tag position relates to the TOF noise, the number of available TOFs and involved tags. In addition, the first term $\bm{H}_{\delta}^T\bm{H}_{\delta}$ in the brackets in (\ref{eq:CRLB_CL}) is related to the TOF measurements, and is the only information utilized in the conventional single tag positioning. The second term relates to the inter-tag distances, and makes the  positioning error $\operatorname{cov}(\Delta\hat{\bm{y}})$ smaller than that of single tag positioning. This theoretically shows the advantage of the ERBL-EDMC method over the single tag positioning method in localization accuracy. 

\subsubsection{CRLB of Position and Attitude Estimation}
The Cram\'er-Rao lower bound (CRLB) is the lower bound of the error variance for the unbiased estimator, which has been used extensively in the literature as a reference for localization accuracy. 
We formulate the CRLBs of $\bm{p}_\mathrm{c} $ and  $ \bm{R} $ for the RBL problem, by extending the CRLBs in  \cite{Chepuri2014RBL_LS,Chen2015RBL_DAC} to the incomplete tag-anchor TOFs case as
\begin{equation}\label{eq:CRLB}
 \mathrm{CRLB}=\bm{C}\left( \bm{C}^T\bm{F}\bm{C}\right)^{-1}\bm{C}^T,
\end{equation}
where $ \bm{F} $ is the FIM for problem (\ref{eq:original_problem}), and $\bm{C}$ is obtained by the SO constraints as will be detailed later of this subsection.

We derive the FIM $ \bm{F} $  as
\begin{equation}
	\bm{F}=\sum_{i\in \mathcal{T}}\sum_{j\in \mathcal{A}_i}\frac{(\bm{l}_{e,i}\otimes\bm{I}_\eta)\Delta\bm{p}_{i\mathcal{A}_i(j)}\Delta\bm{p}_{i\mathcal{A}_i(j)}^T(\bm{l}_{e,i}^T\otimes\bm{I}_\eta)}{\sigma^2 \left\|\bm{p}_\mathrm{c}+\bm{R}\bm{l}_i- \bm{q}_{\mathcal{A}_i(j)}\right\| ^2},
\end{equation}
in which $ \bm{l}_{e,i}=[\bm{l}_i^T,1]^T $ and $\Delta \bm{p}_{i\mathcal{A}_i(j)}=\bm{p}_\mathrm{c}+\bm{R}\bm{l}_i-\bm{q}_{\mathcal{A}_i(j)} $.

Then, we deal with the SO constraints. For 3D case, we express $\bm{R}$ as a combination of column vectors $\bm{\mu}_1$, $\bm{\mu}_2$ and $\bm{\mu}_3$ as $\bm{R}=[\begin{matrix}\bm{\mu}_1&\bm{\mu}_2&\bm{\mu}_3\end{matrix}]  $ and rewrite the SO constraints \eqref{eq:SOcstr} as
\begin{equation}
\begin{aligned}
     \bm{h}(\bm{\chi}) =&\left[
         \bm{\mu}_1^T\bm{\mu}_1-1\quad \bm{\mu}_2^T\bm{\mu}_1\quad\bm{\mu}_3^T\bm{\mu}_1\quad\bm{\mu}_2^T\bm{\mu}_2-1 \right.\\  &\left.\bm{\mu}_3^T\bm{\mu}_2\quad\bm{\mu}_3^T\bm{\mu}_3-1\quad\operatorname{det}(\bm{R})-1 \right]^T =\bm{0}_{7}^T,
\end{aligned}
\end{equation}
where $\bm{\chi}=[\begin{matrix}\bm{\mu}_1^T&\bm{\mu}_2^T&\bm{\mu}_3^T&\bm{p}_\mathrm{c}^T\end{matrix}]^T$. 

 The constraint-related term $ \bm{C} $ for 3D case is then a $12 \times 7 $ matrix as \cite{Chepuri2014RBL_LS}
 \begin{equation}
	\bm{C}=\operatorname{null}(\frac{\partial\bm{h}(\bm{\chi}) }{\partial\bm{\chi}^T}),
\end{equation}
where $\operatorname{null}(\cdot)$ is an orthonormal basis of the null-space of a matrix.

And for 2D case, the constraint-related term $ \bm{C} $ is similarly derived and we have $\bm{C}=\bm{I}$.

We shall use the CRLB in (\ref{eq:CRLB}) as a reference in the simulation of Section \ref{simulation} to evaluate the performance of the proposed method.

\subsection{Computational Complexity}
\begin{table*}[ht]
	\centering
	\begin{threeparttable}	
		\caption{Computational complexity.}
		\label{table:Computational complexity}
		\centering
		\vspace{-0.3cm}
		\begin{tabular}{c c}
			\toprule
			{ Method } & {Computational complexity}\\
			\midrule
			DAC & $O\left( \eta^2M(N-M)+\eta^3+M(N-M)\right) $\\
			SDR &$ O\left( (\eta+1)\left(M(N-M)(\eta+1)^4+M^2(N-M)^2(\eta+1)^2+(\eta+1)^8\right)\right) $\\
			ERBL-EDMC &$ O\left( K_1\eta N^2+\eta^2M(N-M)+K_2\eta^3(N-M)^3+\eta^3+K_3 (\eta+\nu) ^3\right) $\\
			\bottomrule
		\end{tabular}	
	\end{threeparttable}
	\vspace{-0.2cm}
\end{table*}

We study the computational complexity of the proposed method and compare it with DAC \cite{Chen2015RBL_DAC} and SDR \cite{Jiang2019RBL_SDP2} based methods. The complexity is shown in big $ O $ expressions with respect to the number of anchors $  M $, the number of tags $ \left( N-M\right) $ and the localization dimension $ \eta $ (2 or 3). 

For the proposed ERBL-EDMC method in Section \ref{method}, the major computation in Step 1 lies in the construction of the majorization function and updating the EDM for each iteration, which requires $ O(K_1\eta N^2) $ flops \cite{Zhou2018SQREDM}, where $ K_1 $ is the number of iterations to update the EDM. Step 2 requires $ O(\eta^2M(N-M)) $ flops to obtain the coarse tag positions  \cite{Chen2015RBL_DAC} and $ O(K_2 \eta^3(N-M)^3) $ flops to obtain the refined tag positions via the iterative algorithm, and $ K_2 $ is the number of iterations in the refinement step. Specifically, the major computation lies in the matrix
inversion of a $\eta\cdot(N-M)\times \eta\cdot(N-M) $ matrix for each iteration. Step 3 takes $ O(\eta^3) $ flops for SVD to obtain the closed-form solutions \cite{trefethen1997SVD, Chen2015RBL_DAC} and $ O \left(  K_3 (\eta+\nu)^3\right)$ flops to generate the optimal solution iteratively, where $K_3 $ is the number of iterations. The matrix inversion of a $(\eta+\nu) \times (\eta+\nu)$ matrix takes the major computation for each iteration in Step 3.

Table \ref{table:Computational complexity} gives the complexities of the ERBL-EDMC method under the condition that all tag-anchor measurements are available, based on the above analysis. For comparison, we also list in the table the complexities of the DAC and SDR based methods.

\begin{table}[ht]
	\centering
	\begin{threeparttable}	
		\caption{ Computational complexity calculated with typical values.}
		\label{table:Computational complexity cal}
		\vspace{-0.4cm}
\begin{tabular}{@{}cccccc@{}}
\toprule
\multirow{2}{*}{Demension} & \multicolumn{2}{c}{Parameter} & \multicolumn{3}{c}{Methods} \\ 	\cmidrule(r){2-3} \cmidrule(l){4-6} 
                           & $M$             &$N$            & DAC     & SDR  & ERBL-EDMC  \\ \midrule
\multirow{4}{*}{2D}        & 4             & 7             & $O(68)$   &     	$O(26487)$ &	$O(6876)$           \\
                           & 6             & 9             &   $O(98)$&	$O(32805)$&	$O(8180)$           \\
                           & 14            & 17            &     $O(218)$&	$O(77517)$&$	O(16596)$           \\
                           & 16            & 20            &         $O(328)$&	$O(145827)$&$O(27044)$            \\
\multirow{4}{*}{3D}        & 4             & 7             &         $O(147)$&	$O(283648)$&$	O(21975)$           \\
                           & 6             & 9             &         $O(207)$&	$O(301312)$&$	O(23949)$          \\
                           & 14            & 17            &         $O(447)$&	$O(418048)$&	$O(36645)$         \\
                           & 16            & 20            &        $O(667)$&	$O(589824)$&$	O(63483)$         \\ \bottomrule
\end{tabular}
	\end{threeparttable}
	\vspace{-0.2cm}
\end{table}

To compare the computational complexity of the three methods visually, the theoretical estimation of computational complexity with different coordinates dimensions, number of anchors $M$, and number of nodes (including anchors and tags) $N$ is calculated based on the formulas in Table \ref{table:Computational complexity}, and is shown in Table \ref{table:Computational complexity cal}. For Table \ref{table:Computational complexity cal}, the iteration numbers $K_1$, $K_2$ and $K_3$ of the ERBL-EDMC method are set to 20, although they are generally less than 10 in practice. According to this table, it can be found out that the complexity of the proposed method is higher than the DAC method \cite{Chen2015RBL_DAC}, but is lower than the SDR based method in \cite{Jiang2019RBL_SDP2}.

\section{Numerical Simulations}\label{simulation}
In this section, we test the availability, accuracy and computational complexity of the proposed ERBL-EDMC method by numerical simulations. Firstly, numerical scenes without missing measurements are simulated to verify that our method has comparable performance with the DAC method \cite{Chen2015RBL_DAC} and the SDR based method \cite{Jiang2019RBL_SDP2}. Then, a scene with missing TOFs is simulated to examine the performance in the harsh environment with signal blockages. After the discussion on the availability of the new ERBL-EDMC and the existing methods, we present the completed EDM 
result based on our new bound estimation approach and compare it with that from the conventional shortest path method \cite{Zhou2018SQREDM}. The position and attitude estimation errors for the scene with missing TOFs are then given and compared with the DAC and SDR methods.
\subsection{Scene without Missing TOFs}
In this subsection, we construct geometries of anchors for 2D and 3D case with no obstacles to simulate the scenes without missing measurements and compare the performance of the new ERBL-EDMC with the existing methods.

We generate $ \Upsilon =1,000 $ runs for the given anchor geometry. The root mean square error (RMSE) to evaluate the estimates of the position vector $\bm{p}_\mathrm{c}$ and the rotation matrix $\bm{R}$ is calculated by
\begin{align}
\operatorname{RMSE}(*)&=\sqrt{\frac{1}{\Upsilon}\sum_{\upsilon=1}^{\Upsilon}\left\|\hat{*}^{(\upsilon)} -(*)\right\| ^2},
\end{align}
where $ \hat{*}^{(\upsilon)} $ represents the estimates for $\bm{p}_\mathrm{c}$ and $\bm{R}$ at the $ \upsilon$-th run, and $ ({*}) $ is the true value.

\subsubsection{2D Case}
The tag positions in Frame b are set as 
\begin{equation}
	\bm{L}=
	\left[\begin{matrix} 0 &5& 5&0\\
		0& 0 &5&5\\
	\end{matrix}\right],  	
\end{equation} 
with the unit m. We then have the local coordinate $\bm{l}_i=[\bm{L}]_{:,i},\; i \in \mathcal{T}$.
There are $ M = 5 $ anchors placed at $  [40, 50]^T $ m, $  [30, 20]^T$ m, $  [0, 10]^T $ m, $  [-50, -50]^T $ m and $  [-20, -30]^T $ m. The center of the vehicle located at $  [2, 10]^T $ m in Frame g is set as the origin of Frame b. The yaw angle is set to 1.047 rad. We set the standard deviation  $ \sigma $ of the TOF measurement noise varying from 0.1 m to 1 m.  The parameter $\lambda$ in (\ref{eq:problem_CL_linear}) for tag localization is set to 10 in this and the following simulations.

\begin{figure}
	\centering
	\subfloat[RMSE of $\psi$]{
		\includegraphics[width=0.47\linewidth]{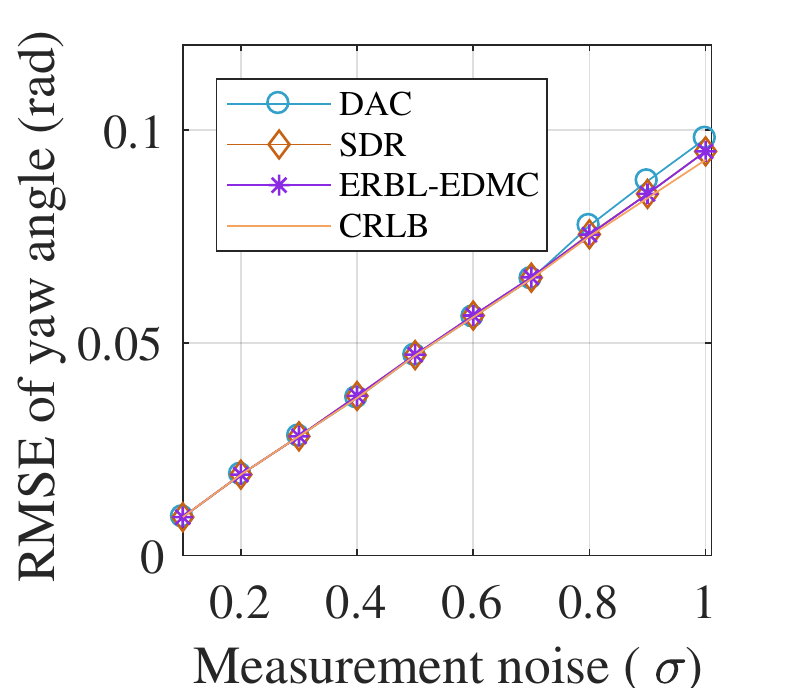}\label{fig:RMSE_2D_yaw}}
	\hspace{0.02\linewidth}
	\subfloat[RMSE of $\bm{p}_\mathrm{c}$]{
		\includegraphics[width=0.47\linewidth]{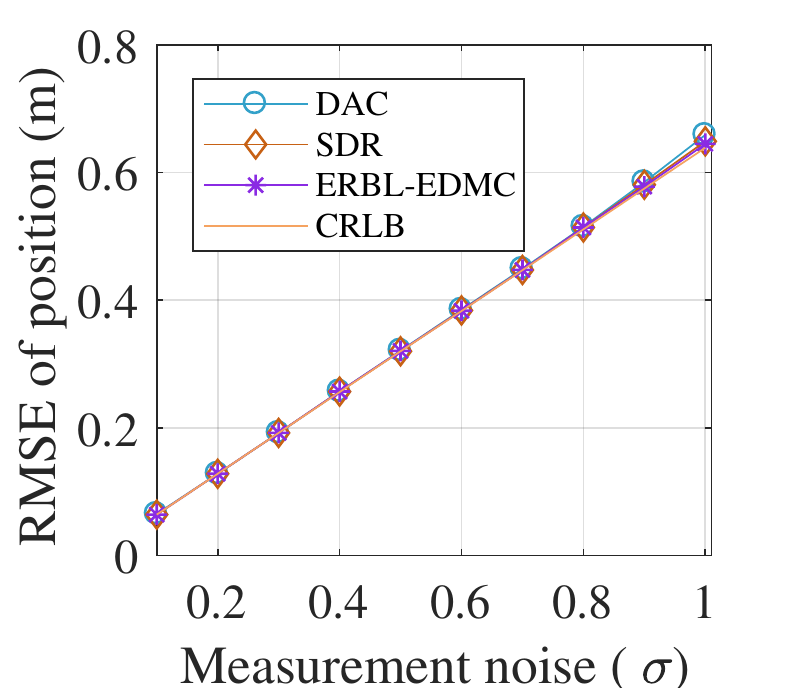}\label{fig:RMSE_2D_p}}
	\vspace{0cm}
	\caption{RMSE vs. measurement noise for 2D case.
		(a) Yaw angle error result. (b) Position error result. The localization errors of the new ERBL-EDMC method are the same as that of the SDR based method. Both the methods have comparable performance close to the CRLB and outperform the DAC method.}
	\label{fig:RMSE_2D}
	\vspace{-0.3cm}
\end{figure}

Fig. \ref{fig:RMSE_2D_yaw} shows the RMSE of the estimation of the yaw angle $ \psi $. The performance of the ERBL-EDMC method is the same as the SDR based method in \cite{Jiang2019RBL_SDP2}. Both methods have comparable performance close to the CRLB and outperform the DAC method. Fig. \ref{fig:RMSE_2D_p} shows the results of the estimated position vector $\bm{p}_\mathrm{c}$. The performance of the  ERBL-EDMC method is better than that of the DAC method, and is the same as that of the SDR based method for this scene. Compared with the DAC method, the ERBL-EDMC method jointly estimate the positions of the tags assisted with the inter-tag distances, which brings more information and thus provides higher accuracy. 

\subsubsection{3D case}

The tag positions in Frame b are set as
\begin{equation}
	\bm{L}=
	\left[\begin{matrix} 1.5 &4.5& 4.5&4.5&3\\
		0& 0&4.5&4.5&3\\
		0&0&0&0&3
	\end{matrix}\right],  	
\end{equation} 
with the unit m. Then, 
the local coordinate $\bm{l}_i$ in (\ref{eq:d_tag}) is the $i$-th column of $\bm{L}$, i.e., $[\bm{L}]_{:,i}$. There are $ M = 6 $ anchors placed at $[-50,-65,-70]^T$ m, $[50,-35,-25]^T$ m, $[-50,5,-5]^T$  m, $[15,-45,-15]^T$ m, $[-15,30,30]^T$  m and $[50,45,55]^T$ m. The center of the vehicle is located at $  [5, 5,2]^T $  m in Frame g. The pitch, roll and yaw angle of the vehicle are set to $-$0.436 rad, 0.349 rad and 0.175 rad, respectively. The standard deviation  $ \sigma $ of the TOF measurement noise is set varying from 0.1  m to 1  m.

\begin{figure}
	\centering
	\subfloat[RMSE of $\bm{R}$]{
		\includegraphics[width=0.47\linewidth]{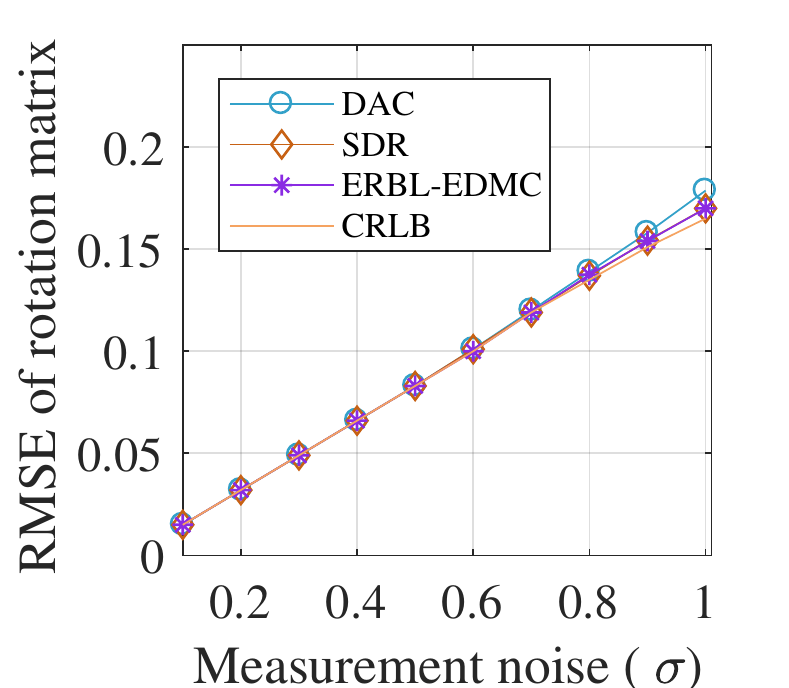}\label{fig:RMSE_R_3D}}
	\hspace{0.02\linewidth}
	\subfloat[RMSE of $\bm{p}_\mathrm{c}$]{
		\includegraphics[width=0.47\linewidth]{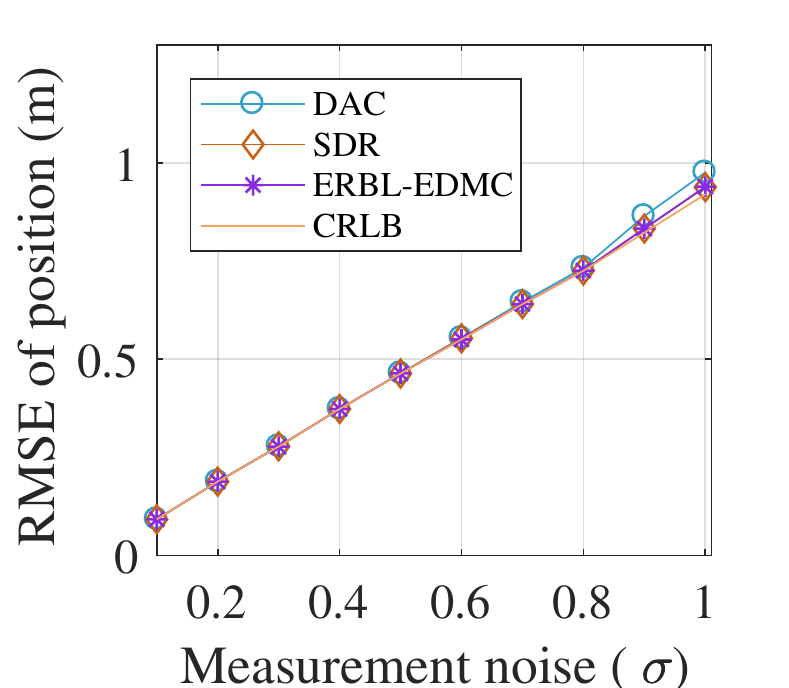}\label{fig:RMSE_p_3D}}
	\vspace{0cm}
	\caption{RMSE vs. measurement noise for 3D case.
		(a) Rotation matrix error result. (b) Position error result. The performance of the new ERBL-EDM method is the same as that of the SDR based method. Both the methods have results close to the CRLB and outperform the DAC method.}
	\label{fig:RMSE_3D}
	\vspace{-0.3cm}
\end{figure}

Fig. \ref{fig:RMSE_R_3D} shows the RMSE of the estimated rotation matrix $\bm{R}$. Similar to the above 2D case, the attitude estimation performance of the new ERBL-EDM method is as good as that of the SDR based method. The estimation accuracies of both the methods are close to the CRLB and are better than the DAC method. Fig. \ref{fig:RMSE_p_3D} is the error result for the position vector $\bm{p}_\mathrm{c}$. The proposed method outperforms the DAC method in positioning accuracy, similar to that of 2D case and consistent with the theoretical analysis.

\begin{table}[ht]
	\centering
	\begin{threeparttable}	
		\caption{Average Run Time}
		\label{table:run times}
		\centering
		\vspace{0.2cm}
		\begin{tabular}{c c c c}
			\toprule
			\multicolumn{2}{c}{2D Case} & \multicolumn{2}{c}{3D Case}\\
	\midrule
			{ Method } & {Run Time (ms)}&{ Method } & {Run Time (ms)}\\
				\cmidrule(r){1-2}	\cmidrule(r){3-4}
			DAC & 1.2 &	DAC & 1.5 \\
			SDR &  404.2&SDR &494.2\\
			ERBL-EDMC & 6.3 &ERBL-EDMC&15.8\\		
\bottomrule
		\end{tabular}	
	\begin{tablenotes}[para,flushleft]
	Note: The run times are the average from 1,000 simulation runs. The complexity of ERBL-EDMC method is higher than that of the DAC method, but is significantly lower than that of the SDR method.
	\end{tablenotes}
	\end{threeparttable}
	\vspace{-0.2cm}
\end{table}

\subsubsection{Complexity}
Table \ref{table:run times} shows the average run times of different methods for 1,000 runs when $\sigma=0.1$ for both 2D and 3D cases. All the methods are implemented using Matlab 2018a on a personal computer with a 3.2-GHz i5-4570 CPU and 32GB RAM. The SDR based method is realized using the Matlab toolbox CVX \cite{Boyd2014cvx} with the solver SeDuMi \cite{Sturm1999solver} with default precision, the same as in \cite{Jiang2019RBL_SDP2}. As presented in Table \ref{table:run times}, the complexity of the new ERBL-EDMC method is higher than that of the DAC method, but is significantly lower than that of the SDR based method, consistent with the complexity analysis in Section \ref{performance}.

\subsection{Scene with Missing TOFs}

For the case with missing TOFs, we construct a scene to simulate an unmanned warehouse as illustrated in Fig. \ref{fig:scene_example}, in which, the goods on the shelves and on the vehicle will block the signals and reduce the number of available TOF measurements. We generate a trajectory for the vehicle and obtain different geometries of the available anchors for each epoch due to the change of relative position between the vehicle, anchors and obstacles.  

Utilizing the simulation data, the availability of our new ERBL-EDMC method is presented. Moreover, we test the effectiveness of the bound estimation approach and examine the accuracy of the ERBL-EDMC method. Finally, iteration counts and computation time at each epoch are also investigated.
 \subsubsection{Scene Setting}
\begin{figure}
	\centering
	\includegraphics[width=0.99\linewidth]{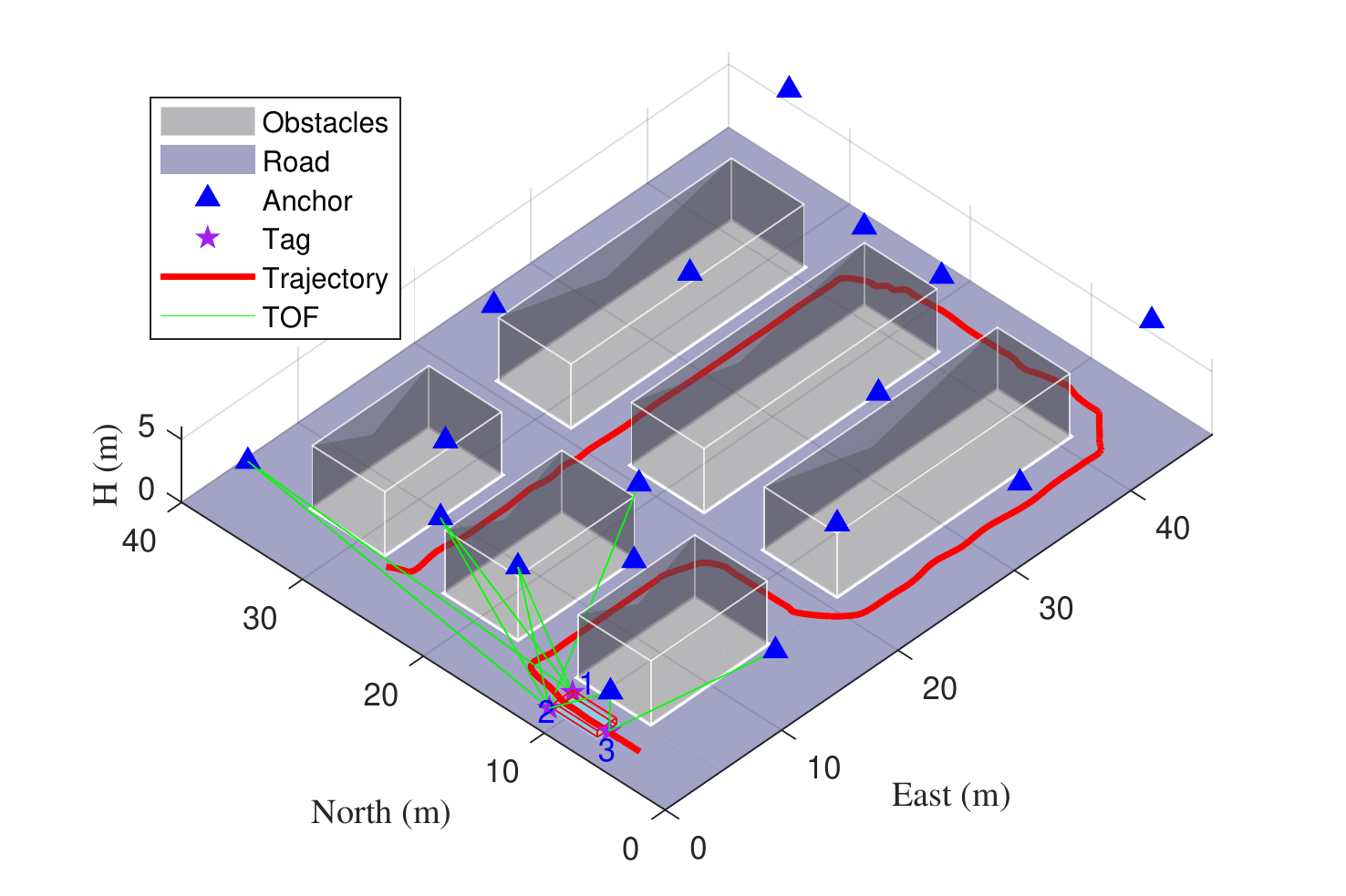}
	\vspace{-0.5cm}
	\caption{Simulation scene for an unmanned warehouse.
		Due to the obstacles in the environment and on the vehicle, the fixed tags on the vehicle can only communicate with a few of the anchors.}
	\label{fig:scene_warehouse_2D}
	\vspace{-0.3cm}
\end{figure}

Fig. \ref{fig:scene_warehouse_2D} shows the simulated unmanned warehouse. There are obstacles such as goods and the shelves represented by the gray cuboids. The vehicle to be located is moving along the aisles between the shelves on the trajectory presented by the red line. Seventeen anchors with known positions in Frame g are placed in the warehouse. Three tags are mounted on the vehicle to obtain more TOFs between the tags and their visible anchors.

\begin{table}[ht]
	\centering
	\begin{threeparttable}	
		\caption{Global coordinates of anchors in the simulated unmanned warehouse scene.}
		\label{table_anchor_scene_3A}
		\centering
		\vspace{0.2cm}
		\begin{tabular}{c  c c  c c c }
			\toprule
			{ Anchor No. } & {East}&{North }&{ Anchor No. } & {East}&{North }\\
			\midrule
			1 & 0.5& 5&10& 20.8&34.2\\
			2 & 4.8& 23.2&11&10&27.8\\
			3 & 0.5&35&12 & 31&27.8\\
			4& 47 & 5&13 & 10&  12.2\\
			5& 41.2& 23.2&14& 31 &  12.2\\
			6 &47&35&15&15.2& 16.8\\
			7 & 10& 0.5&16 & 4.8&16.8\\
			8&  31& 0.5&17 &  41.2& 16.8\\
			9& 20.8&5.8& &  & \\
			\bottomrule		
		\end{tabular}
	\end{threeparttable}
	\vspace{-0.2cm}
\end{table}

The vehicle is 4 m long, 2 m wide and 0.3 m high. The goods with 3.6 m in length, 1.6 m in width and 2.8 m in height are loaded on the vehicle. The obstacles in the environment are 6 m wide and 5 m high. The coordinates of the three tags on the vehicle in Frame b are $[2, 1, -0.15]^T$ m, $[2, -1, -0.15]^T$ m and $[-2, 0, -0.15]^T$ m, respectively.  We conduct 2D positioning in this simulation and set all heights of the anchors as 6 m. The east and north coordinates of the anchors in Frame g are presented in Table \ref{table_anchor_scene_3A} with the unit of meter. 

In this simulation scene, we have $M=17$ anchors and 3 tags. Thus, the number of nodes $N=20$. As shown by the red line in Fig. \ref{fig:scene_warehouse_2D}, the vehicle moves along a trajectory for 300 s. The TOF measurements are updated at a 1-Hz rate. The standard deviations of the TOF noises are set to $\sigma= 0.1$ m, which can be provided by the LiDAR system or well-compensated  UWB-based ranging system \cite{DW2020UWB,pala2020UWBaccurate}. 
We employ the estimation error of the position vector $\bm{p}_\mathrm{c}$ and the yaw angle $\psi$ at each epoch to evaluate the accuracy of our method. The estimation error is calculated by
\begin{align}\label{eq:error_v_m}
	\operatorname{ERROR}{(*)}&=\left\|\hat{*}^{(l)} -*\right\| ,
\end{align}
where $ \hat{*}$ represents the estimates for $\bm{p}_\mathrm{c}$ and $\psi$, and $ ({*}) $ is the true value.

\subsubsection{Availability}
\begin{figure}
	\centering
	\includegraphics[width=0.99\linewidth]{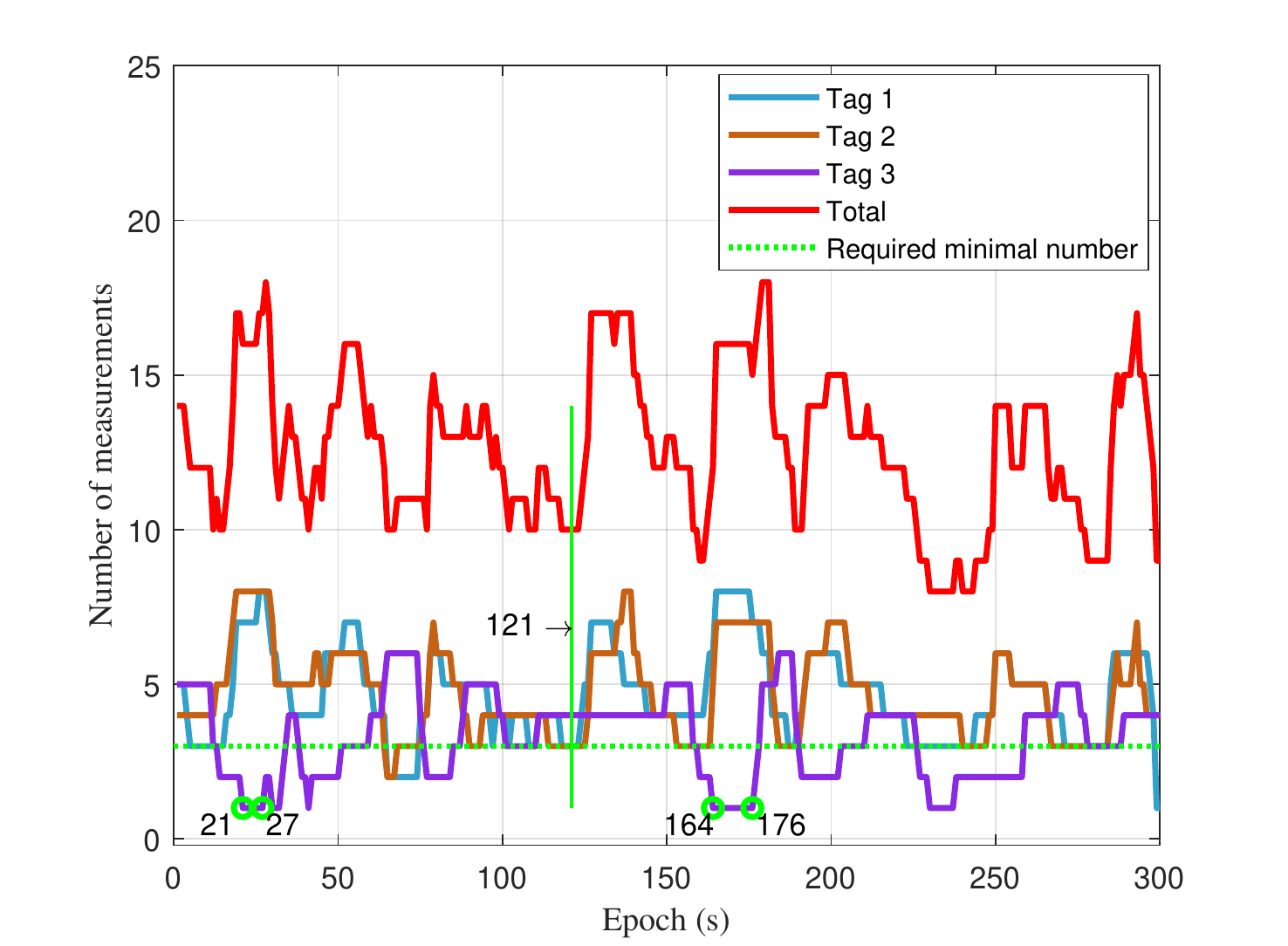}
	\vspace{-0.5 cm}
	\caption{Number of available measurements at different tags. The total number of measurements for all tags is no less than 8 throughout the entire simulation, well beyond the required minimal number for single tag positioning. The availability of ERBL-EDMC method is 100\%. }
	\label{fig:num_anchor}
	\vspace{-0.3cm}
\end{figure}

We record the number of available measurements $M_{i}$ for tag $i,\; i\in\mathcal{T}$ at each epoch, as shown in Fig. \ref{fig:num_anchor}. The green dotted line represents the minimum number of measurements required by single tag positioning. In this scene, we cannot obtain all the tag-anchor TOFs because of the blockages, which lead to incomplete EDMs. Moreover, due to the motion of the vehicle, the geometry and blockages vary from epoch to epoch. 

We can see from Fig. \ref{fig:num_anchor} that the total number of measurements for the all three tags is more than 7 throughout the simulation. Moreover, there are always more than 2 tags that can observe at least one anchor. According to the condition given by \eqref{eq:available}, the availability of the new method is 100\%, i.e., position solution can be obtained throughout the entire trajectory. As for each single tag, the numbers of measurements for tag 1, tag 2 and tag 3 are less than 3 at 12, 3  and 102 epochs, respectively, indicating poor availability for single tag positioning. 

\subsubsection{EDM Completion Error}\label{results_IEDM}
We examine the performance of the proposed bound estimation approach for missing TOFs of EDM completion in Step 1. The Frobenius distances between the completed EDM and the real EDM are adopted as the error of the EDM completion result.

\begin{figure}
	\centering
	\includegraphics[width=0.99\linewidth]{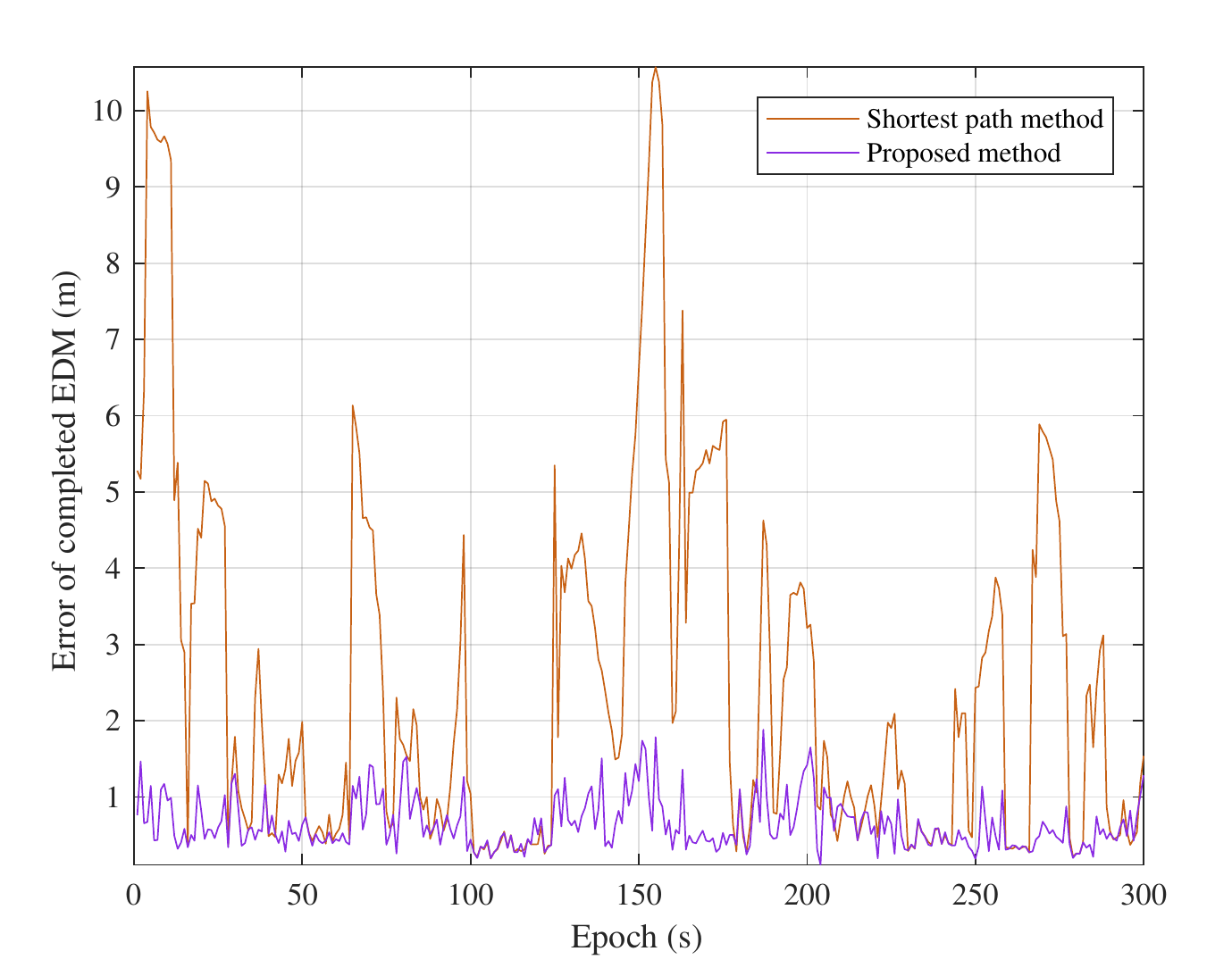}
	\vspace{-0.5 cm}
	\caption{Errors of the completed EDM. The errors from the proposed method are much smaller than those base on the conventional shortest path method throughout this simulation.}
	\label{fig:EDM_error_2D}
	\vspace{-0.3cm}
\end{figure}

Based on (\ref{eq:error_v_m}), the EDM completion errors of Step 1 in the ERBL-EDMC method are presented in Fig. \ref{fig:EDM_error_2D}. The EDM completion errors based on the conventional shortest path method are also depicted for comparison. As shown in the figure, the EDM errors based on the shortest path method are large, unstable and even reach more than 10 m, while those of the new method are smaller than 2 m. It verifies that the bounds of the unspecified elements significantly affect the EDM completion. Moreover, our new bound estimation approach substantially improves the performance of the EDM completion and ensures the accuracy of tag localization.

\subsubsection{Position and Attitude Estimation Error}
\begin{figure}
	\centering
	\subfloat[Error of $\psi$ vs. time]{
		\includegraphics[width=0.99\linewidth]{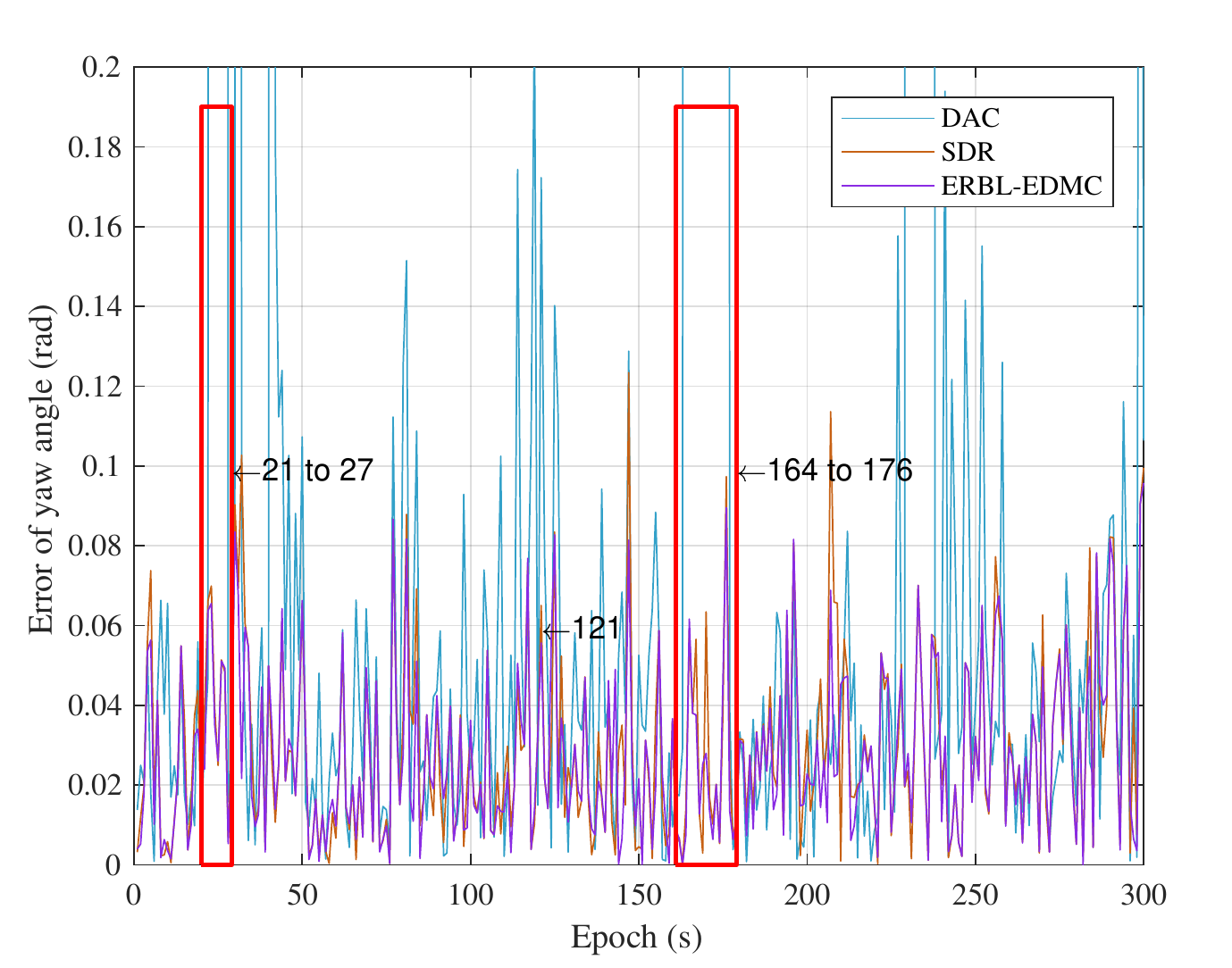}\label{fig:error_2D_p}}
	\vspace{-0.02cm}
	\subfloat[Error of $\bm{p}_\mathrm{c}$ vs. time]{
		\includegraphics[width=0.99\linewidth]{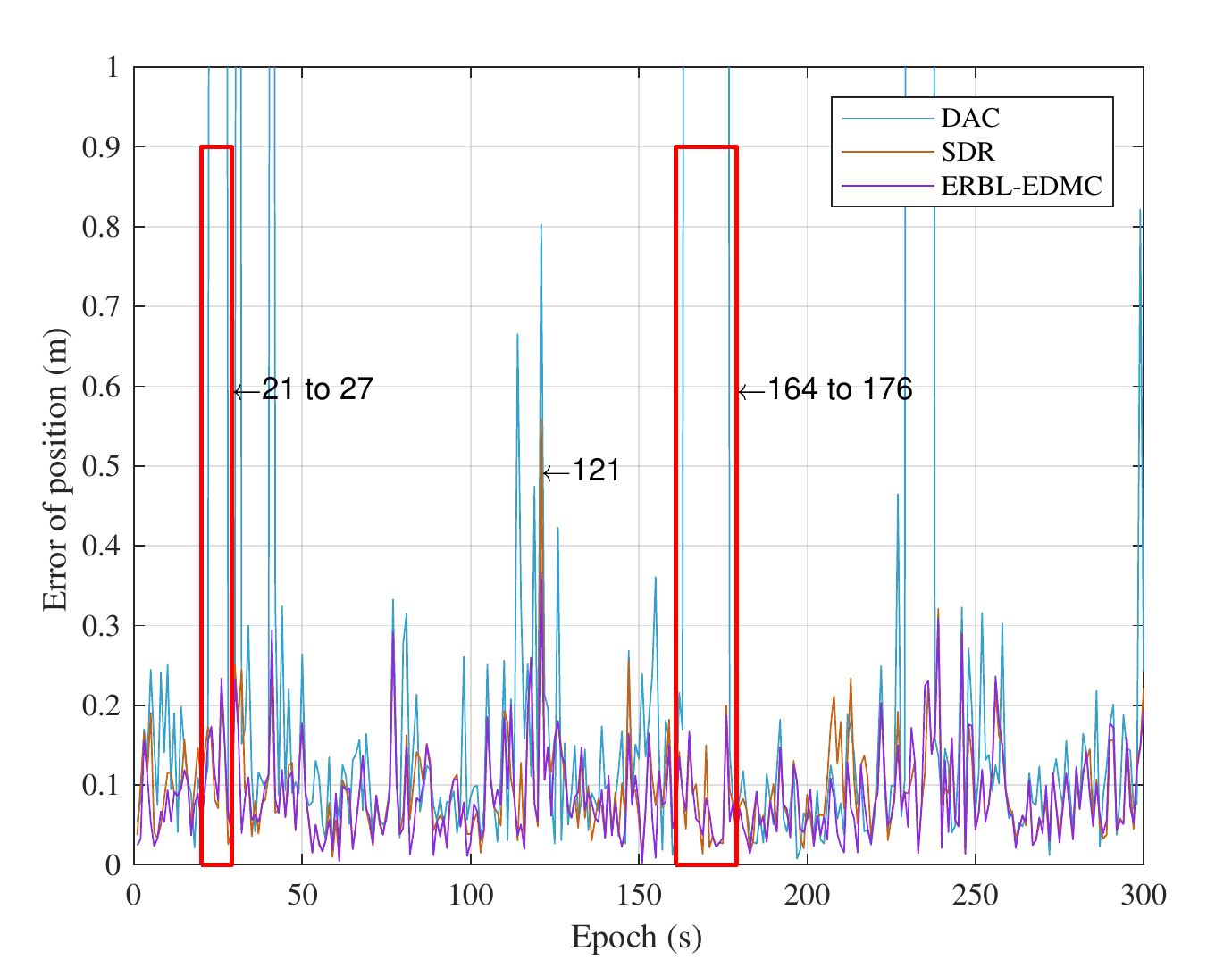}\label{fig:error_2D_yaw}}
	\vspace{0cm}
	\caption{Estimation errors in the unmanned warehouse scene. (a) Yaw angle estimation. (b) Position estimation. The estimation errors of the new ERBL-EDMC method are significantly smaller than those of the DAC method and are slightly smaller than those of the SDR based method. 
}
	\label{fig:error_2D}
	\vspace{-0.3cm}
\end{figure}

Fig. \ref{fig:error_2D_p} and Fig. \ref{fig:error_2D_yaw} show the estimation error from the ERBL-EDMC method throughout the simulation. The errors of the SDR based method and DAC method are also depicted for comparison. The yaw angle error beyond 0.2 rad and the position error beyond 1 m of the DAC method are not presented for better illustration. As shown in the figure, the new method provides accurate and robust position and attitude results for the whole trajectory, in which the yaw angle error and position error are lower than 0.1 rad and 0.4 m, respectively, significantly smaller than those of the DAC method and slightly smaller than those of the SDR based method.

The minimum, maximum and mean errors of $\psi$  and $\bm{p}_\mathrm{c}$ during the simulation are presented in Table \ref{table:statistics}. The results of the DAC and SDR based methods are also given for comparison. The errors of both $\psi$  and $\bm{p}_\mathrm{c}$ for our ERBL-EDMC method are the smallest among the three methods, indicating a better performance with missing TOFs in the harsh environment. In contrast, the DAC method gives very large errors, since at some epochs there are only two tags that can realize single tag positioning and the corresponding estimate of $\bm{p}_\mathrm{c}$ ends up at a wrong position.

\begin{table}[ht]
	\centering
	\begin{threeparttable}	
		\caption{Error statistics in the simulation scene with missing TOFs}
		\label{table:statistics}
		\centering
		\vspace{-0.2cm}
\begin{tabular}{lllllll}
\toprule
\multicolumn{1}{c}{\multirow{2}{*}{Method}} & \multicolumn{3}{c}{$\psi$ (rad)}  & \multicolumn{3}{c}{$\bm{p}_\mathrm{c}$ (m)}\\
\cmidrule(l){2-4} \cmidrule(l){5-7}
\multicolumn{1}{c}{}   & \multicolumn{1}{c}{Min} & \multicolumn{1}{c}{Max} & \multicolumn{1}{c}{Mean} & \multicolumn{1}{c}{Min} & \multicolumn{1}{c}{Max} & \multicolumn{1}{c}{Mean} \\ 
\midrule
  	DAC & 7.9243e-04 &3.250& 0.317&0.008 &4.122 & 0.465   \\
			SDR &  7.0747e-05 &0.123&0.029&0.005&0.558 &0.094 \\
			ERBL-EDMC &1.6988e-05 &0.096&0.028 &0.002 &0.366&0.086 \\
  \bottomrule
\end{tabular}
\begin{tablenotes}[para,flushleft]
	Note: The ERBL-EDMC method has the smallest error among the three methods.
\end{tablenotes}
	\end{threeparttable}
	\vspace{-0.2cm}
\end{table}

To show the superior performance of the new ERBL-EDMC method in detail, we investigate into several epochs with extremely small number of measurements. As illustrated in Figs. \ref{fig:error_2D_p} and \ref{fig:error_2D_yaw}, there are some epochs, e.g.,  21-st to 27-th epochs and  164-th to 176-th epochs as framed by the red rectangles, at which the errors from the DAC method are far greater than that of the other two methods. At these epochs, which are shown in Fig. \ref{fig:num_anchor} by the green circles, tag 3 only have one measurement, such that the single tag positioning cannot work. Utilizing only the positions of tag 1 and tag 2, a wrong vehicle position result is obtained by the DAC method. 
 
\begin{figure}
	\centering
	\includegraphics[width=0.99\linewidth]{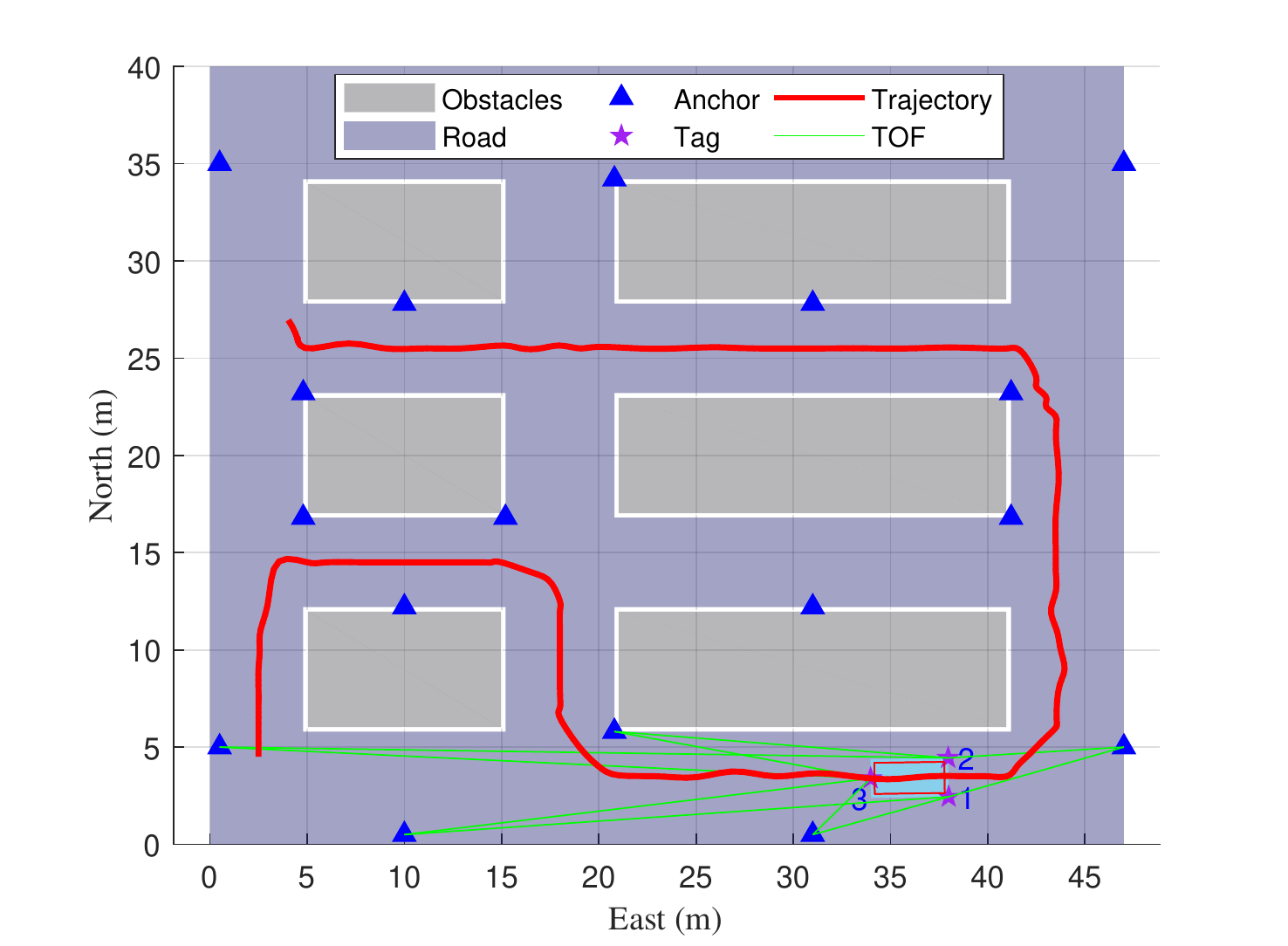}
	\vspace{-0.5 cm}
	\caption{Geometry of visible anchors at 121-st epoch. In this corner area, due to the blockages caused by the obstacles in the surrounding and on the vehicle, the three tags can only obtain 10 TOFs from 5 anchors.}
	\label{fig:epoch}
	\vspace{-0.3cm}
\end{figure}
Moreover, to compare the localization errors between the ERBL-EDMC and the SDR method, we look into the 121-st epoch, at which a small number of tag-anchor measurements are available, as shown in Fig. \ref{fig:num_anchor}. The position of the vehicle and the number of TOFs of each tag are shown in Fig. \ref{fig:epoch}. We can see that due to the blockages caused by the obstacles, the three tags only obtain 10 TOFs from 5 anchors. The errors of the ERBL-EDMC method at this epoch as given by Figs. \ref{fig:error_2D_p} and \ref{fig:error_2D_yaw} are smaller than those of the DAC and SDR based methods, in both of which, the case with incomplete tag-anchor TOFs is not considered, resulting in large errors in such cases.

In addition, we set the standard deviation of the TOF noises as  $\sigma=0.3 $ m, which can represent the UWB ranging error in the non-ideal environment, keep the other settings identical, and conduct another simulation in the same environment. The estimation errors of the yaw angle and position by the new ERBL-EDMC method are lower than 0.3 rad and 0.7 m, respectively. The results are not illustrated in figures to save space. The errors of our ERBL-EDMC method are the smallest among the three methods, consistent with the results of the previous simulation.

\subsubsection{Convergence}
 
\begin{figure}
	\centering
	\includegraphics[width=0.99\linewidth]{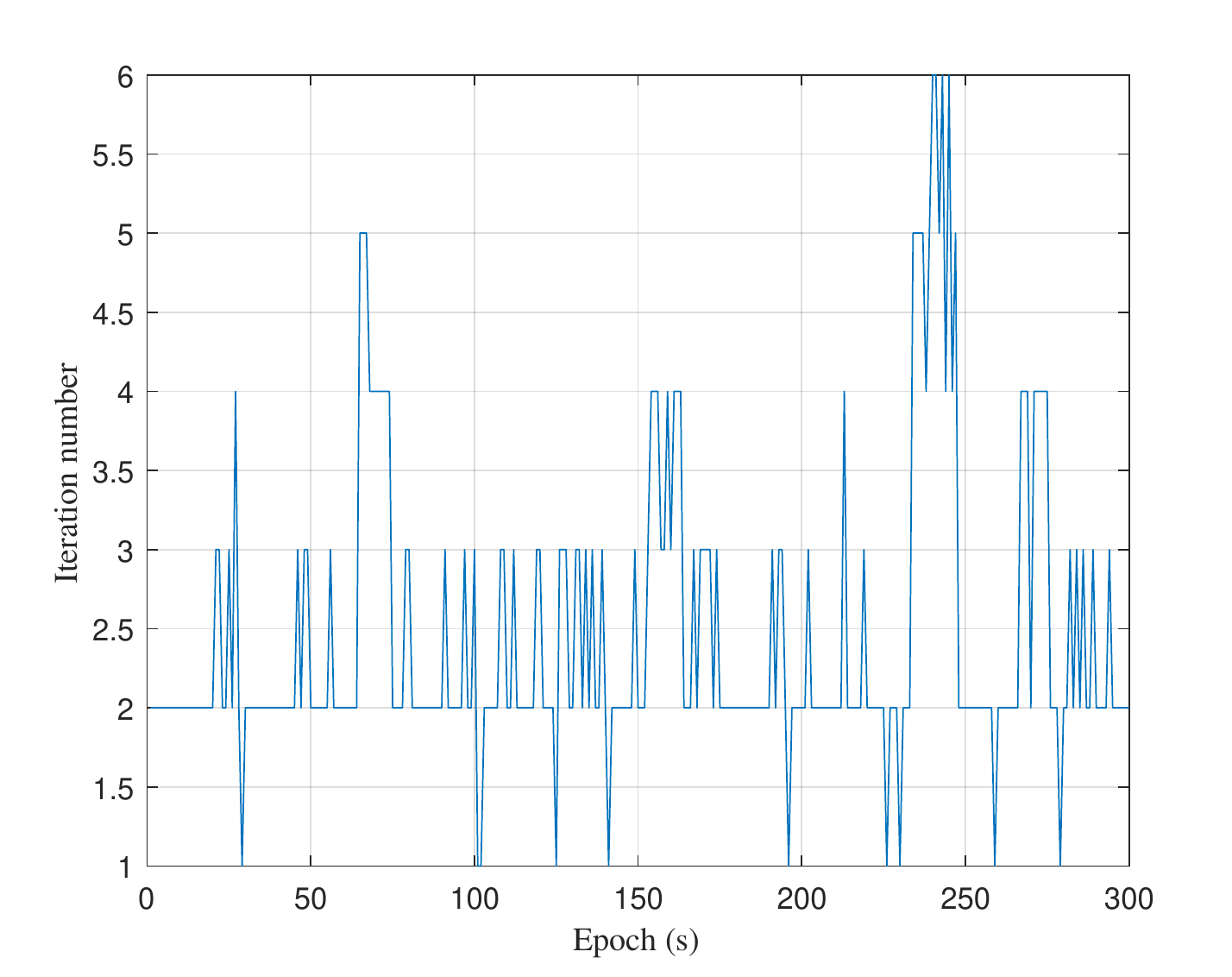}
	\vspace{-0.5 cm}
	\caption{Iteration number vs. epoch. The Position and Attitude Estimation step converges at all epochs within 6 iterations.}
	\label{fig:iteration_epoch}
	\vspace{-0.3cm}
\end{figure}

To analyze the convergence of our method, we look into the number of iterations of the last step at each epoch. As presented in Fig. \ref{fig:iteration_epoch}, the number of iterations at each epoch are no more than 6, which is far less than the preset maximum number of iterations, and shows good convergence performance. It also indicates that the previous steps produce good initial values as input to the iterative algorithm for position and attitude estimation.

\subsubsection{Computation time}
\begin{figure}
	\centering
	\includegraphics[width=0.99\linewidth]{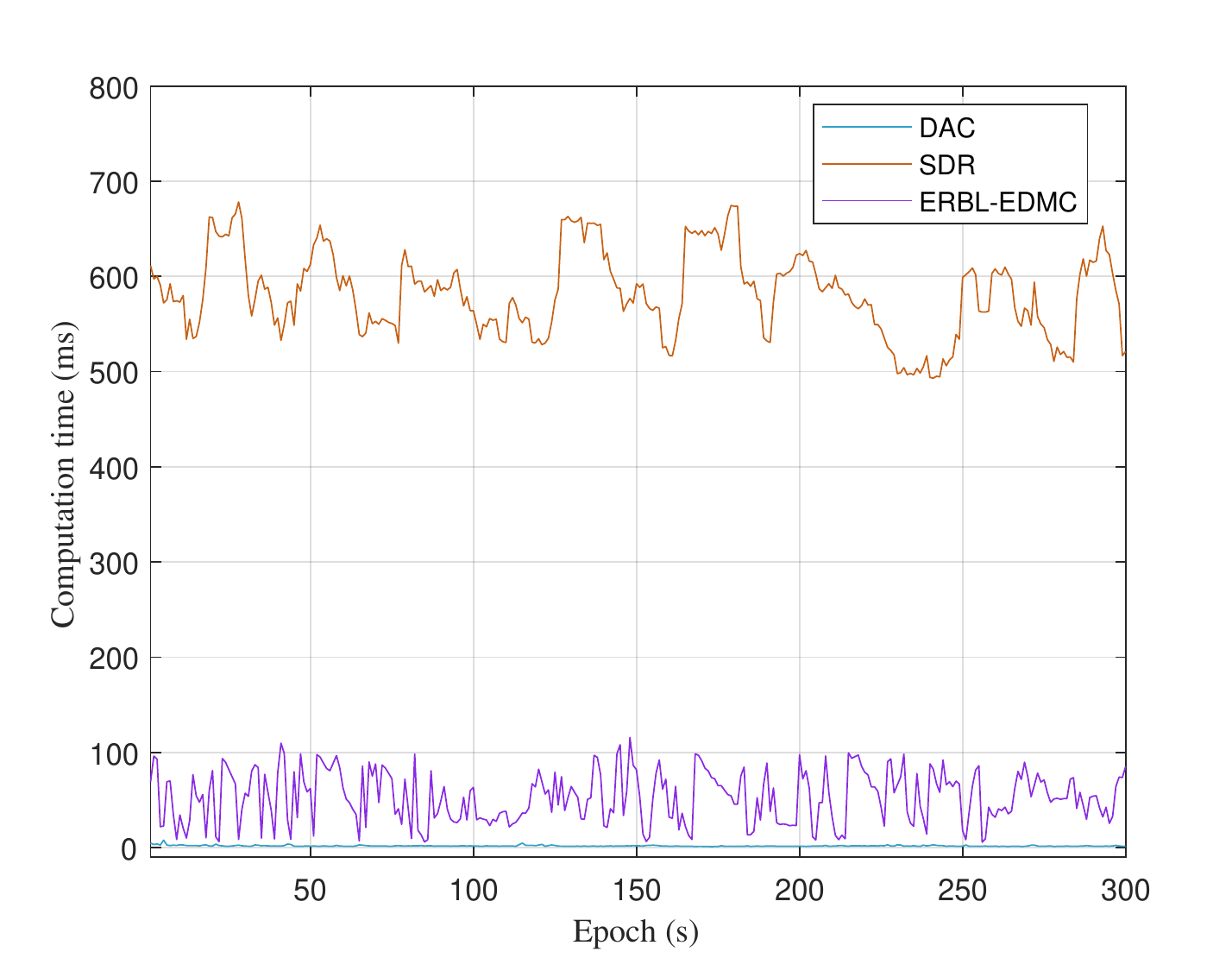}
	\vspace{-0.5 cm}
	\caption{Computation time vs. epoch. The complexity of the new ERBL-EDMC method is significantly lower than that of the SDR based method, and higher than that of the DAC method.}
	\label{fig:time_epoch}
	\vspace{-0.3cm}
\end{figure}

Finally, we investigate the computation time at each epoch and compare it with the existing methods. As shown in Fig. \ref{fig:time_epoch}, our ERBL-EDMC method takes less time than the SDR based method, and more time than the DAC method, which is consistent with theoretical analysis in Section \ref{performance}.

\section{Conclusion}\label{conclusion}
AGV localization based on the radio positioning system in harsh environments with obstacles is challenging. Signal blockages by obstacles reduce the number of available TOF measurements and thus degrade the availability and accuracy of positioning.
To solve the localization problem, we study the radio positioning system that utilizes multiple tags with spatial difference to collect more measurements, and develop a solution with high availability, high accuracy and low complexity for the multi-tag AGV positioning problem with missing TOFs. 

Particularly, by modeling the TOF localization system as an SN, we propose an efficient RBL method based on EDM completion, namely ERBL-EDMC method. To reliably complete the original measured EDM with noisy and unspecified elements, we develop a new bound estimation approach for the RBL problem by using the known local tag positions and the statistics of the available TOF measurements. Furthermore, by using the multi-tag TOFs assisted with inter-tag distances and a coarse estimation based on the completed EDM, accurate global tag positions are jointly obtained, ensuring a good initialization and thus the optimal results for the final position and attitude estimation. Theoretical analysis and numerical results show that the proposed method has optimal positioning accuracy and low computational complexity, and outperforms the state-of-the-art methods.


\ifCLASSOPTIONcaptionsoff
    \newpage
\fi




\bibliographystyle{IEEEtran}
\bibliography{IEEEabrv,paper_an}



\end{document}